\documentclass[10pt,preprint]{aastex}
\usepackage{graphics,graphicx}
\shortauthors{GWINN AND JOHNSON}
\shorttitle{}
\begin{document}
\input epsf

\title{Noise and Signal for Spectra of Intermittent Noiselike Emission}

\author{C. R. Gwinn, M.D. Johnson}
\affil{Department of Physics, University of California, Santa Barbara, California 93106, USA}
\email{cgwinn@physics.ucsb.edu,michaeltdh@physics.ucsb.edu} 

\vskip 1 truein
\begin{abstract}
We show that intermittency of noiselike emission,
after propagation through a scattering medium,
affects the distribution of noise in the observed correlation function.
Intermittency also affects
correlation of noise among channels of the spectrum,
but leaves
the average spectrum, average correlation function, and distribution of noise among channels of the spectrum unchanged.
Pulsars are examples of such sources: intermittent and affected by interstellar propagation.
We assume that the source emits Gaussian white noise, modulated by a time-envelope.
Propagation convolves the resulting time series with an impulse-response function that represents effects of dispersion, scattering, and absorption.
We assume that this propagation kernel is shorter than the time for an observer to accumulate a single spectrum.
We show that rapidly-varying intermittent emission tends to concentrate noise near the central lag of the correlation function.
We derive mathematical expressions for this effect, in terms of the time envelope and the propagation kernel.
We present examples, discuss effects of background noise, and compare our results with observations.

\end{abstract}

\keywords{methods: data analysis -- techniques}

\section{INTRODUCTION}\label{sec:introduction}

The observed spectrum of a source with intrinsically intermittent emission,
including propagation effects,
is the focus of this paper.
Propagation effects can often be described as a convolution, in time, with an impulse-response function or ``propagation kernel''.
This tends to smear out any intrinsic variation of the source.
In particular, if the emission from the source is noiselike, but modulated in time, 
then propagation will mix emission at different times, so that the observer perceives noiselike emission of nearly constant intensity.
However,
we find that imprints of intermittency remain, under certain simple assumptions.

Our work is motivated by observations of pulsars at meter and decimeter wavelengths.
Pulsars show a rich variety of time variability, on timescales as short as nanoseconds, as we discuss below.
Propagation broadens these rapid intrinsic variations with propagation kernels, with typical time spans of microseconds to milliseconds.  
The forms of these kernels vary over longer timescales, of seconds or more. 
Our goal is to distinguish between intrinsic and propagation effects, as simply and with as few approximations as possible.

A vast variety of propagation effects can be described as a convolution of the emission, in time, with a propagation kernel.
Such a convolution introduces correlations among samples, which an observer can measure.
We suppose that the source emits Gaussian-distributed noise, with time-varying amplitude.
Most astrophysical sources emit noise of this type,
and it provides the standard model for pulsar emission.
Because our assumptions are relatively broadly based, we anticipate that our results may have wider applicability.
An observer measures spectra or correlation functions from the time series, 
and estimates an average, regarded as the determinstic part, or signal; and variations about that average, usually regarded as noise.
We suppose that the observer accumulates data for a single spectrum over a timescale short compared with that for the propagation kernel to vary,
and makes statistical comparisons among spectra. 

Using our model approximations, 
we calculate the averages and noise for spectra and correlation functions.
Among the various factors that contribute to noise,
we focus in particular on self-noise, produced by the noiselike emission of the source. 
We find that intermittency introduces correlations of the noise among spectral channels in the spectrum,
and alters the distribution of noise in the correlation function.
The average spectrum and correlation function of the source are the same as those of a non-varying source, with the same propagation kernel.
Intermittency of the source affects only noise, but affects it in a way that can be calculated.
Thus, noise carries information about the source that is lacking in averages.

\subsection{Outline}\label{sec:outline}

In this paper, we discuss the effect of intermittent emission of a noiselike source
on the spectrum and correlation function, in the presence of propagation effects.
In the following \S\ref{sec:theory_background}, 
we introduce treatment of propagation as a convolution,
and quantify the approximations used for this treatment.
We discuss noiselike emission and relate it to the noise and averages of the observables.
We describe observations of the pulsar B0834$+$06 as an example, and evaluate our approximations.

In \S\ref{sec:math_treatment}, we present formal calculations of the averages, or signal; and of noise, for the spectrum and correlation function.
This section forms the core of this paper.
We introduce notation and fundamentals of noise, and introduce the time-envelope and the 
propagation kernel, and the observed time series (\S\ref{sec:math_intro}).   
We transform to the frequency domain of the power spectrum, 
and present expressions for the average spectrum, and for the variances and covariances of noise of spectral channels (\S\ref{sec:spectral_domain}).
We then transform back to the time-lag domain of the correlation function
and present expressions for the average correlation function and noise in the correlation function (\S\ref{sec:lag_domain}).
All of these quantities depend on the propagation kernel.
However, 
only noise in the correlation function, and correlation of noise between spectral channels, depend on the
time-envelope of emission at the source.
We present examples of spectrum and correlation function
for simple cases of source intermittency, and simple propagation kernels, in \S\ref{sec:examples_redux}.

In \S\ref{sec:extensions} we discuss comparisons and extensions of our mathematical theory.
We present connections to the amplitude-modulated noise theory of \citet{Ric75}, 
describe effects of background sky or receiver noise,
and compare our results with observations.
In \S\ref{summary} we summarize our results.

\section{THEORETICAL BACKGROUND}\label{sec:theory_background}

\subsection{Propagation as a Convolution}\label{sec:convolution}

A convolution in time describes effects of wave propagation through a medium with spatially-varying index of refraction, static in time. 
Multiplication of the spectrum by a frequency-dependent complex scalar provides an equivalent description,
via the convolution theorem for Fourier transforms \citep{Bra00}.
Applications include 
cell-phone communication, seismology, ocean sound propagation, radar and microwave circuits, adaptive optics, and interstellar radio-wave propagation
\citep{Sim01,Mou00,Bul85,Zio85,She09,Mon87,goo00,Gwi98}. 
For electromagnetic waves, this fact stems from the superposition principle.
A localized assemblage of charges and currents with 4-vector current density $J^{\mu}(t, \vec x )$,
gives rise to a 4-vector potential $A^{\mu}(t, \vec x )$.
If current density varies at a single frequency $\omega$,
then $A^{\mu}$ is a solution to the Helmholtz equation \citep{Lan75}. 
In this equation, a spatially-varying wavenumber 
describes a time-independent, but spatially non-uniform, index of refraction \citep{Ers07}. 
The solution for $A^{\mu}(\omega, \vec x)$ for an arbitrary arrangement of refracting material
exists and is unique under quite general assumptions.
Helmholtz equations of this form arise for propagation of many kinds of waves, and 
a host of techniques provides means of solving them \citep[][and references therein]{Fla79,Zio85,goo00,Ers07}.

For a particular frequency, source configuration, and arrangement of refracting material,
the electric field at the observer equals that at a fiducial point near the source, times a complex propagation factor $\tilde g$.
The propagation factor can depend on frequency.
For time-independent refraction, $\tilde g(\omega)$  is independent of the phase or amplitude of the electric field at the source. 
More generally, a matrix $\tilde g(\omega)$ relates 
different source polarization states, and different source locations, to the electric field in different observed polarization states and observer locations.
This is the S-matrix or scattering matrix, originally 
developed to describe quantum-mechanical scattering \citep{Cus90}, and then adapted for use in electromagnetic propagation
\citep{Mon87,Mou00,Sim01}.
A superposition of sources leads to a superposition of received electric fields;
under some circumstances, such as a sufficiently small source, the details of the source distribution are unimportant 
\citep{Gwi98,goo00,Sim01}.

Multiplication by the scalar $\tilde g(\omega)$ in the frequency domain 
becomes convolution by its Fourier transform $g(\kappa)$ in the time domain, where $\kappa$ is the time lag 
(\citealp{Bra00}; see also eq.\ \ref{eq:convolution_theorem} below).
Indeed, for any linear time-invariant system, of arbitrary dimensionality and complication,
one time series input and one time series output are always related by convolution with such a kernel \citep{Rab75,Opp97}. 
The kernel $g(\kappa)$ is known as the ``propagation kernel'' or the ``impulse-response function''.
Its width is the typical scattering time $\tau$, expressed in terms of time lag $\kappa$.

\subsection{Spectrum and Dynamic Spectrum}\label{sec:dynamic_spectrum}

\subsubsection{One spectrum}

This paper is concerned with statistics of spectra and correlation functions.
In this paper, we use the terms ``one power spectrum'',  ``one realization of the power spectrum'', or ``an estimate of the power spectrum''
for the square modulus of the discrete Fourier transform of a time interval of data, of $N_{\nu}$ samples of duration $\delta t$ each. 
The Fourier transform of the power spectrum is the autocorrelation function.
Interferometry combines two time series: the product of the Fourier transform of one with the conjugated Fourier transform of the other
is one realization of the cross-power spectrum.
The Fourier transform of the cross-power spectrum is the cross-correlation function \citep{tms01,Per89}.
In this paper, we use the term ``spectrum'' to refer to either the power spectrum or the cross-power spectrum.
Sometimes the power spectrum is known as the ``autocorrelation spectrum'' or simply the ``auto spectrum'';
and the cross-power spectrum is sometimes known as the ``cross spectrum''.
The Fourier transform has both discrete and integral forms; \citet{tms01} describe the two forms and their relations.
This paper is concerned with finite spans of sampled data, and hence we use the discrete Fourier transform.   
In some formal contexts, one realization of the power spectrum is known as the ``periodogram'', whereas the ``spectrum" is the 
integral Fourier transform of a continuous time series \citep{Bro06};
however, we follow traditional usage here. 

We call an average of spectra over an ensemble of statistically-identical spectra
with different realizations of noise ``the ensemble-average spectrum''
or simply the ``average spectrum''.
We denote specifically averages over time, frequency, or propagation kernel, when we invoke them.
Effects of intermittency or propagation are not included in our statistical ensemble: only those of noise.
Averaging the autocorrelation function over an ensemble of realizations of noise yields the ``average autocorrelation function''.

This paper deals with scattering that is 
approximately time-independent, over the time interval for accumulation of a single spectrum.
f the medium remains time-independent for many accumulation times,
consecutive spectra are statistically identical, and the observer can average them together to increase signal-to-noise ratio.
We discuss these considerations in detail in \S\ref{sec:math_treatment}.

\subsubsection{Dynamic spectrum}

Over longer times, the propagation kernel varies and the spectra differ.
This paper is not directly concerned with this regime, 
although we briefly discuss analysis of data with such variations in \S\ref{sec:extensions}.
A simple way of dealing with time series that vary with both frequency and time is 
the dynamic spectrum or, more formally, short-time Fourier transform (STFT) \citep{Bra00,Che02,Boa03}.
The observer divides a longer time series 
into segments of $N_{\nu}$ elements each, and forms one spectrum from each segment.
Individual spectra can be averaged together for periods shorter than the timescale for the propagation kernel to change.
The time series of such spectra is the dynamic spectrum, a two-dimensional function of time and frequency.

The 2D Fourier transform of the dynamic spectrum is the autocorrelation function in the lag-rate domain.
The Fourier transform of the spectrum is the autocorrelation function, as noted above.
The Fourier transform along the time axis converts the time series of autocorrelation functions to frequency, known as ``rate".  
The square modulus of the resulting lag-rate autocorrelation function is the ``secondary spectrum''.
The secondary spectrum is particularly well-suited to study of the largest delays produced by interstellar scattering,
which is observed to take the form of ``scintillation arcs'' in this domain \citep{Hil03}.
Theoretical description of scattering is quite advanced, and makes extensive use of the dynamic spectrum
and secondary spectrum to infer the statistics of scattering material, and distribution in particular cases \citep{Ric77,Wal04,Cor06}.

\subsection{Approximations}\label{sec:assumptions}

In this section we relate our fundamental approximations to the properties of source, scattering material,
and instrumentation for observations.
The assumption of time-independent refraction is an approximation for all the applications mentioned above; but in many cases the approximation is excellent. 
These are fundamentally instrumental 
approximations in that their accuracy depends on the time period the observer samples.
If the time for accumulation of a single realization of the spectrum is short, the medium can be treated as time-independent.
We also require
that the time for accumulation of a spectrum is longer than the time span of the propagation kernel.
This is likewise an instrumental constraint.
Finally, we assume that the source is noiselike:
more precisely, the source emits a random electric field, drawn from a Gaussian distribution at each instant and uncorrelated in time.
The variance of the Gaussian distribution gives the amplitude of the source.
This paper deals specifically with the effects of intermittency on short timescales, where the variance of the distribution changes
during the time for accumulation of one spectrum.

\subsubsection{Time variation of propagation kernel}\label{sec:time_variations}

Propagation can be treated as time-independent when the time to accumulate a spectrum 
is shorter than the characteristic time for the propagation medium
to change significantly, at the retarded time.
Propagation includes dispersion and scattering.
Dispersion changes only slowly with observing epoch \citep{Bac93},
far too slowly for variations to affect a single spectrum.
Scattering changes much more rapidly.
For most media, the propagation kernel for scattering has a relatively rapid rise, with a much slower fall \citep{wil72,bol03a}.
Within that envelope, phase and amplitude vary rapidly as the electric field along different paths cancels or reinforces. 
The propagation kernel changes with time, with changes in the scattering medium or locations of source and observer.

Motion of the material in the propagation medium produces quasistatic evolution of the solution to the Helmholtz equation.
A slow variation of the propagation factor $\tilde g$ with time represents the effect \citep{Zio85}.
For small-angle forward scattering by angle $\theta$ and speed $v$,
the local rate of change of phase is about $\theta \nu v/c$,
corresponding to a time for change by about 1 radian of order $t_d \approx c/(\theta \nu v)$.
For observations of galactic objects at distances up to a few kiloparsecs, we expect scattering material to move at 
$v\approx 10$ to $100\ {\rm km\ sec}^{-1}$,
and $\theta$ some fraction of the observed scattering angle. 
In many cases, motion of the source sets the timescale for scintillation;
the scattering material is assumed ``frozen'' as it moves across the line of sight 
\citep{goo00}. This assumption is the foundation of velocity measurement of pulsars, via scintillation \citep{Gup95}.
In this case, the expression for $t_d$ is similar, with $v$ now the speed of the source perpendicular to the line of sight,
and with factors of order 1.
This assumption is also fundamental to most theories of material responsible for scattering \citep[see][]{Wal04}.
For galactic sources observed at meter or shorter wavelengths, typical timescales $t_d$ are seconds or longer.
The typical accumulation time for an individual spectrum $N_{\nu}\delta t$ is about a millisecond.  Thus, the approximation is quite good. 
We present a specific example in \S\ref{sec:example} below.

\subsubsection{Doppler shift}\label{sec:doppler}

The approximation of time-independent refraction holds when a single spectrum has spectral resolution insufficient to detect 
Doppler shifts from motions of the source or medium.
In this case, a propagation factor $\tilde g (\omega)$ describes scattering accurately.
For small-angle forward scattering, the change in frequency from Doppler shift is
$\delta\nu \approx \theta \nu v/c$.
This expression is the same as that in the previous section, and indeed the net effect of Doppler shifts can be expressed as a time derivative of 
$\tilde g (\omega)$ \citep{Zio85}. 
For typical galactic sources,
$\Delta \nu /\nu \approx 10^{-11}$\ for scintillation arcs, and much less for the typical scattering \citep{Wal04}.
The frequency resolution of a typical spectrum, $1/N_{\nu} \delta t$, is typically a few kilohertz.
Thus, these Doppler shifts are not detectable in individual spectra, at observing frequencies less than 100 GHz.
Effects of the Doppler shifts do appear in dynamic spectra,
as quasi-static evolution of the propagation kernel in time.
We present an example in \S\ref{sec:example}.

\subsubsection{Time span of the propagation kernel}\label{sec:kernel_span}

In our calculations, we require that the time span of the propagation kernel,
$\tau$, is shorter than the time for accumulation of a single spectrum, $N_{\nu} \delta t$.
The kernel includes effects of both scattering and dispersion.
For scattering time $\tau_d$ and scintillation bandwidth $\Delta\nu_d=1/2\pi \tau_d$,
this becomes the requirement that the spectral resolution be finer than $\Delta\nu_d$.
This requirement is mildly violated for some observations of scattering at the largest $\tau$, as we discuss in \S\ref{sec:example}.

In practice, the timescale for dispersion $\tau_{DM}$ can be longer than the timescale for the spectrum,
as we discuss in \S\ref{sec:example}.
This does not present a fundamental obstacle, because the contribution of dispersion to the propagation kernel can be inverted.
Dispersion correction via multiplication by the conjugate of the expected phase of $\tilde g(\omega)$, 
is precise but computationally intensive and is usually reserved for the most interesting cases \citep{Han71}.
Incoherent dedispersion alleviates effects of dispersion \citep{Vou02}, but is not a pure deconvolution and can be expected to leave some artifacts.
Ideal measurements would involve coherent dedispersion, or long accumulation times for single spectra, or both.

\subsection{Noiselike Emission}\label{sec:noise_def}

We assume that the emission from the source is noiselike.
Specifically, we assume that the source emits Gaussian white noise:
the electric field at each instant is random and drawn from a Gaussian distribution, and
the field at different instants is uncorrelated \citep{Pap91}.
Many sources, including astrophysical sources under almost all conditions, emit such radiation, because they comprise many superposed, independent radiators  \citep{Dic46,Evans72}. The distribution of electric field is then Gaussian because of the central limit theorem.
We do not assume that the emission of the source is stationary; in other words,
we suppose that the variance of the underlying Gaussian distribution may vary with time.

The noise in an observable $V$ is the departure of a given measurement from its statistical average:
\begin{equation}
\delta V = V - \langle V \rangle_n.
\end{equation}
This average is calculated 
over an ensemble of statistically-identical realizations of self-noise and background noise, denoted by the subscripted angular brackets $\langle ... \rangle_n$. 
``Self-noise'' or ``source noise''
is the contribution of the noiselike emission of the source to noise of observables \citep{kul89,ana91,Gwi04,Gwi06}.
We quantify this effect in \S\ref{sec:math_spectrum_noise} below, and describe observations in \citet{Gwi11}.
Background noise includes Gaussian noise from sky, ground and antenna emission, and noise in the receiving system.
The ensemble does not include variations in the flux density of the source or of the propagation kernel. 
We treat these effects with the functions $f$ and $g$, defined below. 
Consequently,
in this paper, we do not approximate the statistical ensemble by averages over time or frequency, 
and we do not include effects of changes in time envelope or propagation kernel on periods longer than the period to form a single spectrum. 
The variance of the observable $V$ quantifies that noise: 
\begin{equation}
\langle \delta V^2 \rangle_n = 
\delta V^2  \equiv
\langle V^2 \rangle_n - \langle V \rangle_n^2 .
\end{equation}
If the distribution of noise is Gaussian, then $V$ is 
completely characterized by its mean, the signal; and its variance, the noise in $V$ \citep{Pap91}.  

\subsection{Example Observations}\label{sec:example}

In this section we describe observations, and evaluate the approximations discussed in \S\ref{sec:assumptions}.
We consider observations of scintillations of the pulsar B0834$+$06.
\citet{Gwi11} report single-dish observations at Arecibo and very-long baseline interferometry observations on
the Arecibo-Jodrell Bank baseline; \citet{Bri10} report very-long baseline interferometry observations using a network of antennas, including
the Arecibo-Green Bank baseline.
Both observed at frequency $\nu \approx 320$\ MHz,
with observing bandwidth of $\Delta \nu \approx 8\ {\rm MHz}$.
Both used incoherent dedispersion \citep{Vou02}.
Instrumental parameters important for the approximations are as follows:
The time for accumulation 
of a single spectrum was 
$N_{\nu} \delta t = 0.33\ {\rm msec}$ for single-dish and 0.128 msec for VLBI observations of 
Gwinn et al.;
and $8\ {\rm msec}$ for Brisken et al.
All of the observations averaged spectra together in time to increase signal-to-noise ratio:
Gwinn\ et\ al. 
averaged over $\Delta t_S=10$\ sec,
and Brisken\ et\ al.  
over 5 pulses, or $\Delta t_S =6.4$\ sec.
All observations gated synchronously with the pulsar pulse;
thus, only a few individual  spectra of the many in these time intervals contributed to averages.

Parameters of the source were measured in these and previous observations.
Dispersion broadens a short pulse by $\tau_{DM}\approx 23\ {\rm msec}$ at $\nu =320$\ MHz \citep{Lor04}.
Scintillation produces strong variations of intensity across the spectrum,
characterized by a typical scintillation bandwidth $\Delta\nu_d = 0.57$\ MHz \citep{Gwi11}.
The characteristic timescale of the propagation kernel is $\tau_d = 1/2\pi \Delta\nu_d = 0.3\ \mu{\rm sec}$.
The time span of the kernel is a few times greater.
The scintillation timescale is $t_d= 290$\ sec \citep{Gwi11}. This is the timescale for variation
of the propagation kernel.
Note that these observational and instrumental parameters meet with the requirements in \S\ref{sec:assumptions}:
duration of the propagation kernel is less than accumulation time for a spectrum, 
which is less than the averaging time, which is less than time for variation of the propagation kernel:
$\tau_d < N_{\nu}\delta t <\Delta t_S < t_d$.
Note that dispersion violates the requirement that the propagation kernel is shorter than the
accumulation time for a spectrum: $\tau_{DM}>N_{\nu}\delta t$.  
However, incoherent dedispersion mitigates the violation.
The inferred angular broadening is about $\theta_d \approx 1\ {\rm mas}$,
so scattering is small-angle forward scattering.

Fine-scale structure appears in the dynamic spectrum of pulsar B0834$+$06, corresponding to a scintillation arc \citep{Hil03}.
The arc represents about 3\% of the spectral power, so it is a small, but interesting, effect of propagation \citep{Gwi11}.
The arc extends to delay as great as $\tau_a = 1.2\ {\rm msec}$, and to rates as high as $\omega_a = 2\pi \times 50\ {\rm mHz}$ \citep{Bri10}.
This rate corresponds to a timescale for variation of the propagation kernel of $t_a \approx 20\ {\rm sec}$ .
For the arcs, the approximations hold well for the observations of Brisken et al.: $\tau_d < N_{\nu}\delta t <\Delta t_s< t_a$.
Gwinn et al. miss the longest-timescale part of the propagation kernel for the arcs, but this contains relatively little spectral power.
The inferred angular deflections range up to $\theta_a\approx 30\ {\rm mas}$, still small-angle forward scattering.

\section{MATHEMATICAL THEORY}\label{sec:math_treatment}

In this section, we calculate average spectrum and correlation function, and noise, using our approximations of \S\ref{sec:assumptions}.
Table\ \ref{symbol_table} summarizes the symbols we use to describe emission, propagation, and observation.
Table\ \ref{equation_table} summarizes our mathematical results.

\subsection{Definitions and Notation}\label{sec:math_intro}

\subsubsection{Noise}\label{sec:noise}

We model source emission as Gaussian white noise of constant amplitude. 
This noise is multiplied by a time-envelope function to model
intermittent emission.
This noiselike, but nonstationary emission is then subjected to spectral variation, which expresses the effects of propagation.
An instrument converts an electric field to a complex time series via a Hilbert transform \citep{Bra00}
and then converts a limited band of frequencies to a baseband time series \citep{Lor04}.
The observer forms products of samples of the baseband time-series to form spectra or correlation functions.

The spectra may vary greatly with time; as an example, intrinsic emission from pulsars varies greatly with time, 
and interstellar propagation imposes time-varying spectral modulation, by convolution with the propagation kernel.
It is essential that the observer not normalize the spectra, or otherwise impose a time-varying gain. 
This would change the variances of the underlying contributions to noise, independently of the variations introduced by source emission
and the particular realization of propagation, so that underlying contributions would be impossible to disentangle.
Similarly, we define normalizations of theoretical quantities in this section. 

Noiselike emission has an electric field $e_j$.
Here, the index $j$ represents time.
Each sample $e_j$ is drawn from a 
complex Gaussian distribution with zero mean and unit variance. 
We suppose that the underlying noiselike emission is ``white'':
different samples are uncorrelated  \citep{Pap91}.
Then, 
\begin{equation}
\langle e_j e_k^* \rangle_n = \delta _{j k}.
\label{eq:noise_delta}
\end{equation}
Here, $\delta_{j k}$ is the Kronecker delta symbol,
which is 1 if $j=k$ and 0 otherwise.
Because the noiselike emission has no intrinsic phase, 
\begin{equation}
\langle e_j e_k \rangle_n = \langle e_j^* e_k^* \rangle_n =0,
\label{eq:noise_like_zero}
\end{equation}
whether $j=k$ or not \citep{Gwi06}.
We leave open the possibility of different noiselike sources:
for example source and background noise, or background noise for two different antennas;
we denote these different electric fields with superscripts: $e_j^{{x}}$, $e_k^{{y}}$, and so on.
Different ``noises'' are uncorrelated:
\begin{equation}
\langle e_j^{{x}} e_k^{{y} *} \rangle_n = \delta _{j k}\, \delta _{x y}.
\label{eq:noise_delta_xy} 
\end{equation}
We discuss the consequences for background noise in \S\ref{sec:with_background_noise}.

\subsubsection{Flux Density}\label{sec:flux_density}

The intensity of the source (or, equivalently, the flux density) is proportional to the mean square of the electric field.
For a noiselike electric field, intensity is the variance \citep{Dic46}.
A noiselike source of constant flux density has stationary electric field.
Variations in the time envelope and in the propagation kernel can both change the flux density.

To describe sources of arbitrary flux density, we add a prefactor of the average flux density, $I_s$.
Because we consider only pairwise combinations of sources of noise in this mathematical section, 
we can omit the prefactor $I_s$ in this mathematical section without loss of generality.
We normalize the noise using Eq.\ \ref{eq:noise_delta_xy} above, the time envelope using Eq.\ \ref{eq:f_normalization},
and the propagation kernel using Eq.\ \ref{eq:g_normalization};
effects of any flux density variations are subsumed into $I_s$.
We 
reintroduce $I_s$ in \S\ref{sec:extensions} for comparison with other work and observations, where multiple components of noise 
with different variances must be taken into account.
This parameter also
takes into account the possibility of variations in average flux density among spectra because of changes in the time-envelope or propagation kernel.

\subsubsection{Time Envelope}\label{sec:time_envelope}

The emission may be intermittent: 
the electric field may be modulated in time.
In this case, the electric field is a white, Gaussian random variable; but it is not stationary \citep{Pap91}.
We accommodate this by scaling $e_j$ by the emission envelope $f_j$.
This function expresses a change in the standard deviation of the emission,
without changing its noiselike character.
Multiplication by $f_j$ in the time domain is equivalent to convolution with its Fourier transform in the spectral domain;
we make much use of this fact below.
We demand that the time-envelope of emission, $f_j$, not change the average intensity:
\begin{equation}
\sum_{j=0}^{N_{\nu}-1} f_{j} f_{j} = N_{\nu} . \label{eq:f_normalization}
\end{equation}
Without this normalization, variations in source emission will change the means and variances of the correlation function
and spectrum, making impossible the calculation of signal and noise (\S\ref{sec:math_spectrum} and \S\ref{sec:lag_domain}).  
We assume that $f_j$ is real and positive:
a change in phase of $f_j$ has no effect on the statistics of the underlying noiselike emission.
Noise from different sources can have different time-envelopes $f^{x}$, $f^{y}$: 
noise from a source may vary, while noise from a receiver remains constant.
For later convenience, we introduce the products of time-envelopes:
\begin{equation}
\beta_j^{{x}} = f_{j}^{{x}} f_{j}^{{x}}, \quad  \quad \beta_j^{{y}} = f_{j}^{{y}} f_{j}^{{y}}, \quad  \quad \beta_j^{{xy}} = f_{j}^{{x}} f_{j}^{{y}} 
\label{eq:beta_def}
\end{equation}
If the time envelopes are the same, $f_{j}^{{x}}=f_{j}^{{y}}$, all the $\beta$'s are the same:
$\beta_{n}^{{x}} = \beta_{n}^{{y}}=\beta_{n}^{{xy}}$.

\subsubsection{Propagation Kernel}\label{sec:propagation}

After emission, propagation affects the intermittent, noiselike emission by convolution in the time domain, as discussed in
\S\ref{sec:convolution} above.
We parametrize this effect by convolving with the kernel $g_k$.
This kernel includes all propagation effects: dispersion, scattering, and even spectrally-varying absorption.
It can also include the spectral response of the observer's instrument.
Convolution with $g_k$ is equivalent to multiplication by its Fourier transform in the spectral domain.
Similarly to $f_j$, we demand that the propagation kernel not affect average intensity: 
\begin{equation}
\sum_{j=0}^{N_{\nu}-1}\, g_j^{{x}}\, g_j^{{x}*} = 1 . \label{eq:g_normalization}
\end{equation}
Like Eq.\ \ref{eq:f_normalization}, this normalization allows calculation of signal and noise in \S\ref{sec:math_spectrum} and \S\ref{sec:lag_domain}.  
As argued below, products of $g$'s are the average correlation function or spectrum, $\alpha$ or $\rho$.
For a stationary white-noise source, without time or spectral variations or propagation effects, $f_j^{{x}}=1$, and $g_j^{{x}} = \delta_{j\,0}$.

\subsubsection{Time Domain and Wrap Assumption}\label{sec:correlation}

To make a measurement, an observer forms products 
of two, possibly distinct, electric fields $x_{\ell}$ and $y_{\ell}$. 
To obtain these fields, the underlying noiselike emission from the source $e$ is multiplied by $f$ and convolved with $g$:
\begin{equation}
x_{\ell} = \sum_{j=0}^{N_{\nu}-1} g_{j}^{{x}}\, f_{\ell-j}^{{x}} e_{\ell-j}^{{x}} \label{eq:x_ell_def}
\end{equation}
and similarly for $y$.
Here, the span of the observation is $N_{\nu}$ samples.
The mean intensity is $\langle x x^*\rangle_n$, and mean interferometric visibility is $\langle x y^*\rangle_n$.
More generally, the time series  
can be correlated with different lags and Fourier transformed to form a spectrum;
or they can be Fourier transformed and then multiplied together to form a spectrum \citep{tms01}.

For convenience, and for compatibility with the circularity of the discrete Fourier transform in \S\ref{eq:fourier_def}, we make the ``wrap'' assumption 
throughout this paper \citep{Gwi06}: 
\begin{equation}
X_{i+N_{\nu}}=X_i , \label{eq:wrap_assumption}
\end{equation}
for any quantity $X$.
This simplifies calculations, and accurately describes an ``FX'' correlator \citep{Chi84}.
Some ``XF'' correlators violate this assumption, but effects on noise are calculable and relatively minor \citep{Gwi11b}.
Moreover, we form circular convolutions and correlations \citep{Rab75,Opp97}. 
Padding of the $N_{\nu}$ samples with zeros recovers the common case of ignoring the wrap. 

The different fields $x_{\ell}$ and $y_{\ell}$ might arise at different observing stations as in interferometry, or from different sources (such as source noise and background noise, or different spatial regions of one source).
For interferometry of a single pointlike source, the underlying noiselike emission $e$ and time-envelopes $f$ are identical for two stations
$x$ and $y$, but propagation $g$ may be different.
We will retain the identifying superscripts in the results throughout this section, so as to extend the model to include 
multiple components, for discussion in \S\ref{sec:with_background_noise}.

\subsection{Frequency Domain}\label{sec:spectral_domain}

\subsubsection{Fourier transform, Parseval's theorem,  convolution theorem}\label{sec:fourier_convention}

The Fourier transform relates the electric field in the time domain $X_{\ell}$, with that in
the frequency domain $\tilde X_k$.  We adopt the normalization convention for the Fourier transform
and the inverse transform:
\begin{eqnarray}
\tilde X_{k} &=& \sum_{\ell={-N_{\nu}/2}}^{N_{\nu}/2-1} e^{i {{2\pi}\over{N_{\nu}}} k \ell} X_{\ell} \label{eq:fourier_def}
\\
X_{\ell} &=& {{1}\over{N_{\nu}}} \sum_{k={-N_{\nu}/2}}^{N_{\nu}/2-1} e^{-i {{2\pi}\over{N_{\nu}}} k \ell} \tilde X_{k} \nonumber
\end{eqnarray}
Here, $N_{\nu}$ is the number of spectral channels.
This is the discrete form of the Fourier transform \citep{tms01}.  

Parseval's Theorem states that power is conserved by the Fourier transform.
Thus,
\begin{equation}
\sum_{k=0}^{N_{\nu}-1} | \tilde X_k |^2 = N_{\nu} \sum_{\ell=0}^{N_{\nu}-1} |X_\ell |^2 . \label{eq:parseval}
\end{equation}
Power is conserved separately for signal and noise;
thus, this equation  
holds for the square of statistically-averaged $\langle V\rangle_n$ and for noise power $\delta V^2$
as well as for individual time series, spectra and correlation functions.

The convolution theorem, sometimes known as the Faltung theorem, states that the Fourier transform of the convolution of two functions
is the product of their Fourier transforms. Thus,
\begin{eqnarray}
\tilde V_k \tilde U_k &=& \sum_{m=0}^{N_{\nu}-1} e^{i {{2\pi}\over{N_{\nu}}} k m} \sum_{\ell=0}^{N_{\nu}-1} V_{\ell} U_{m-\ell} \label{eq:convolution_theorem}
\end{eqnarray}
for any functions $V$ and $U$.
Using the fact that the Fourier transform of $X_{-\ell}^*$ is $\tilde X_{k}^*$
(see Eq.\ \ref{eq:fourier_def}),
\begin{eqnarray}
\tilde V_k^* \tilde U_k &=& 
\sum_{m=0}^{N_{\nu}-1} e^{i {{2\pi}\over{N_{\nu}}} k m} \sum_{j=0}^{N_{\nu}-1} V_{j}^* U_{m+j} 
\label{eq:convolution_theorem_variation}
\end{eqnarray}
For the inverse transform, the convolution theorem takes the form:
\begin{eqnarray}
V_{m} U_{m} &=& {{1}\over{N_{\nu}^2}} \sum_{k=0}^{N_{\nu}-1} e^{-i {{2\pi}\over{N_{\nu}}} k m} \sum_{j=0}^{N_{\nu}-1} \tilde V_{j} \tilde U_{k-j} \label{eq:convolution_theorem_tolag}
\end{eqnarray}

\subsubsection{Data series in the spectral domain}\label{sec:data_series_in_spectral_domain}

In the spectral domain, we deal with the Fourier transforms of the time series. These Fourier transforms are the spectral series
$\tilde x_k$, $\tilde y_k$. 
These spectral series, in turn, can be described in terms of the Fourier transforms of 
the underlying noise, 
$\tilde e_k^{{x}}$, $\tilde e_k^{{y}}$;
the time-envelopes 
$\tilde f_k^{{x}}$, $\tilde f_k^{{y}}$;
and the propagation kernels
$\tilde g_k^{{x}}$, $\tilde g_k^{{y}}$.
Using the definition of $x_{\ell}$ and the convolution theorem Eq.\ \ref{eq:convolution_theorem},
we find
\begin{eqnarray}
\tilde x_k = \tilde g_k \sum_{j=0}^{N_{\nu}-1} \tilde f_j \tilde e_{k-j}
\end{eqnarray}
Because the Fourier transform of white noise is white noise \citep{Pap91},
$\langle \tilde e_k^{{x}} \tilde e_\ell^{{y} *}\rangle_n = \delta_{k \ell}\cdot\delta_{e}$. 
Here, we introduce the notation $\delta_e$ to express a value of 1 if $e_x$ and $e_y$ are identical, and 0 if they are different. 
Because of Parseval's Theorem,
$\sum_{k} \tilde f_k^{{x}} \tilde f_k^{{x} *} =N_{\nu}$,
and $\sum_{k} \tilde g_k^{{x}} \tilde g_k^{{x} *} =1$, and similarly for the quantities depending on $y$. 
Because the time-envelope functions are real, $\tilde f_k^{{x}} = \tilde f_{-k}^{{x} *}$.

\subsection{Spectrum}\label{sec:math_spectrum}

\subsubsection{Average Spectrum}\label{sec:avg_spectrum}

When two spectral series are multiplied together, $\tilde x_k\,\tilde y_k^*$,
their product is
one realization of the cross-power spectrum, as in an FX correlator \citep{Chi84,tms01}.
This is
\begin{eqnarray} 
\tilde r_k &\equiv& {{1}\over{N_{\nu}}} \tilde x_k\tilde y_k^* ={{1}\over{N_{\nu}}} \sum_{\ell,m=0}^{N_{\nu}-1} 
\tilde g_{k}^{{x}}\, \tilde f_{k-\ell}^{{x}}\,  \tilde e_{\ell}^{{x}}\; \tilde g_{k}^{{y} *}\, \tilde f_{k-m}^{{y} *}\, 
\tilde e_{m}^{{y} *}.  \label{eq:single_realization_spectrum}
\end{eqnarray}
If a single series is multiplied by itself, its square modulus is one realization of the power spectrum:  
\begin{equation}
\tilde a_k^{{x}} \equiv {{1}\over{N_{\nu}}} \tilde x_k\tilde x_k^* , \label{eq:single_realization_ac_spectrum}
\end{equation}
The expression for $\tilde a_k^{{x}}$ in terms of $\tilde g^{{x}}$, $\tilde f^{{x}}$, and $\tilde e^{{x}}$ is identical to Eq.\ \ref{eq:single_realization_spectrum}, 
but with $y\rightarrow x$.
The expression for $\tilde a_k^{{y}}$ is analogous, but with $x\rightarrow y$.
The factor of $1/N_{\nu}$ is required for self-consistent normalization.

The statistically-averaged spectrum $\tilde \rho_k$ (or $\tilde \alpha_k$ for the power spectrum) 
is the deterministic part of this realization.
We calculate this as an average over noise, while holding the time-envelope and propagation kernel constant.
We again denote the average with subscripted angular brackets $\langle ...\rangle_n$:
\begin{eqnarray}
\tilde \rho_k \equiv \langle \tilde r_k\rangle_n &=& \tilde g_{k}^{{x}}\, \tilde g_{k}^{{y} *}\, {{1}\over{N_{\nu}}} \sum_{\ell,m=0}^{N_{\nu}-1} 
\tilde f_{k-\ell}^{{x}}\, \tilde f_{k-m}^{{y} *}\, 
\left\langle \tilde e_{\ell}^{{x}} \tilde e_{m}^{{y} *} \right\rangle_n  .\label{eq:tilde_rho_as_g} \\
&=& \tilde g_{k}^{{x}}\, \tilde g_{k}^{{y} *}\cdot \delta_{e} \nonumber 
\end{eqnarray}
where we have made use of the fact that different samples of noise are uncorrelated, Eq.\ \ref{eq:noise_delta_xy}.  
Again, the symbol $\delta_e$ indicates a factor of 1 if the noise series $e_x$ and $e_y$ are identical, and 0 if they are different. 
This expression defines the ensemble-average cross-power spectrum $\tilde \rho_k$.
Interestingly, this average depends only on the propagation kernel $g$.
If the underlying noise $e$ is different for $x$ and $y$ 
(as for combination of source noise with background noise) then $\tilde \rho_k = 0$,
as the symbol $\delta_{e}$ indicates. 
The autocorrelation functions also depends only on the propagation kernel:
\begin{eqnarray}
\tilde \alpha_k^{{x}} \equiv \langle \tilde a_k^{{x}}\rangle_n &=& \tilde g_{k}^{{x}}\, \tilde g_{k}^{{x} *}, \label{eq:tilde_alpha_as_g}
\end{eqnarray}
and likewise for $\alpha_k^{{y}}$.
Because of the normalization of the propagation factor $\tilde g$ discussed in \S\ref{sec:data_series_in_spectral_domain},
the normalization of $\tilde\alpha$ is $\sum_{k=0}^{N_{\nu}-1} \tilde \alpha_k = 1$.
Because of Parseval's Theorem, for our normalization of the Fourier transform, 
the flux density of the source averaged over time is equal to the flux density summed over the spectrum.
In our calculations we have set both to 1 (see \S\ref{sec:flux_density}).
Although these results are presented in the context of spectral variations imposed by scattering,
they are valid for a noiselike source with any sort of imposed spectral variations, including 
line absorption. 
The notation here is identical to that of \citet{Gwi06}.

\subsubsection{Noise in the spectrum}\label{sec:math_spectrum_noise}

Noise in the spectrum is the departure of a realization of the spectrum from the average.
The variance in a spectral channel characterizes noise in that channel.
More generally, we can calculate the covariance of noise in two (possibly different) channels as
\begin{eqnarray}
\delta (\tilde r_k \tilde r_{\ell}^* ) 
&=& \langle \tilde r_{k} \tilde r_{\ell}^* \rangle_n -  \langle \tilde r_{k}\rangle_n\langle \tilde r_{\ell}^* \rangle_n \label{eq:spect_noise_rkrkc}\\
&=& {{1}\over{N_{\nu}^2}} \sum_{p,q,r,s=0}^{N_{\nu}-1} 
\tilde g_k^{{x}}\, \tilde f_{k-p}^{{x}}\, \tilde g_{\ell}^{{x} *}\, \tilde f_{\ell-r}^{{x} *}\, \left\langle\tilde e_p^{{x}}\, \tilde e_r^{{x} *}\right\rangle_n
\tilde g_{\ell}^{{y}}\, \tilde f_{{\ell}-s}^{{y}}\, \tilde g_{k}^{{y} *}\, \tilde f_{k-q}^{{y} *}\, \left\langle\tilde e_s^{{y}}\, \tilde e_q^{{y} *}\right\rangle_n  \nonumber \\
&=& \tilde \alpha_k^{{x}}\, \tilde \alpha_{\ell}^{{y} *}\, 
{{1}\over{N_{\nu}}} \sum_{p=0}^{N_{\nu}-1} \tilde f_{k-p}^{{x}}\, \tilde f_{p-{\ell}}^{{x} }\, 
{{1}\over{N_{\nu}}} \sum_{q=0}^{N_{\nu}-1} \tilde f_{q-\ell}^{{y} *}\, \tilde f_{k-q}^{{y} *}\nonumber \\
&=& \tilde \alpha_k^{{x}}\, \tilde \alpha_{\ell}^{{y} *}\, \tilde \beta_{k-\ell}^{{x}}\, \tilde \beta_{k-\ell}^{{y} *}. \nonumber 
\end{eqnarray}
We have used the fact that $\tilde f_k = - \tilde f_{-k}^*$,
as well as the ever-crucial fact that samples of the emitted electric field are uncorrelated, Eqs.\ \ref{eq:noise_delta} and \ref{eq:noise_delta_xy}.
Here, $\tilde \beta_{k}$ is the convolution of the time-envelope function with itself, in the spectral domain:
\begin{equation}
\tilde \beta_{k}^{{x}} = {{1}\over{N_{\nu}}} \sum_{p=0}^{N_{\nu}-1}  \tilde f_{p}^{{x}}\, \tilde f_{k-p}^{{x} }
\label{eq:beta_def_convolve}
\end{equation}
and similarly for $\tilde \beta_{k}^{{y}}$. The convolution theorem shows that $\tilde \beta$ is the Fourier transform of $\beta$, Eq.\ \ref{eq:beta_def}.
Because of Parseval's Theorem, and our normalization of $f$ in the 
time domain,
$\tilde \beta_{0}^{{x}} = \tilde \beta_{0}^{{y}} = 1$.
This expression remains unchanged if the underlying noiselike time series $\tilde e_k^{{x}}$ and $\tilde e_k^{{y}}$ are uncorrelated:
then the mean correlation of $x$ and $y$ is $\tilde \rho_k = 0$ as noted in \S\ref{sec:avg_spectrum} above, but 
the noise is the same, given by Eq.\ \ref{eq:spect_noise_rkrkc}.
For the power spectrum of $x_{\ell}$, the same expression gives the noise, with of course $\tilde r_k \rightarrow \tilde a_k^{{x}}$ on the left-hand side of the equation, 
and with $\tilde \alpha_k$ and $\tilde \beta_k$ all for the same station and propagation kernel on the right: ${y}\rightarrow {x}$. 
The analogous construction yields noise for $\tilde a_k^{{y}}$.

For a single spectral channel $k$, Eq.\ \ref{eq:spect_noise_rkrkc} simplifies greatly, and becomes:
\begin{eqnarray}
\delta (\tilde r_k \tilde r_k^* ) = \langle \tilde r_{k} \tilde r_{k}^* \rangle_n -  \langle \tilde r_{k}\rangle_n\langle \tilde r_{k}^* \rangle_n &=& \tilde \alpha_k^{{x}}\, \tilde \alpha_k^{{y} *} \label{eq:spect_noise}
\end{eqnarray}
This equation implies that intermittent emission, as expressed by the time-envelope function, does not change the noise in an individual spectral channel at all.
In fact, this equation states that the signal-to-noise ratio in a single spectral channel is 1: 
$\delta (\tilde r_k \tilde r_k^* )/|\rho_{k}|^2 = 1$ \citep[see][]{Bro06}.
This is consistent with the usual radiometer equation \citep{Dic46}, 
if one observes that the number of samples per spectral channel is 1 .
If the observer averages together $N_t$ independent samples of the spectrum, then the signal-to-noise ratio increases, as the noise falls proportionately to $1/N_t$,
or more slowly if amplitude variations are present \citep{Gwi11}.

In general, the time-envelope function can affect the covariance of noise between spectral channels, as Eq.\ \ref{eq:spect_noise_rkrkc} suggests.
Indeed, the normalized covariance of noise between two channels depends only on the time-envelope function 
and the separation of channels $(k-\ell)$:
\begin{equation}
{{\delta (\tilde r_k \tilde r_{\ell}^* )}\over{\sqrt{\delta (\tilde r_k \tilde r_k^* ) \delta (\tilde r_{\ell} \tilde r_{\ell}^* )}}}
= \tilde \beta_{k-\ell}^{{x}}\, \tilde \beta_{k-\ell}^{{y} *} .
\label{eq:normalized_noise_covariance}
\end{equation}
The normalized covariance does not depend on the propagation kernel or the particular channels $k$ or $\ell$.
For a source emitting white noise of constant intensity, $\tilde \beta_{k}=\delta_{k\, 0}$, and
the covariance of noise in different channels is zero.  For a source emitting a single emission spike during the integration,
$\tilde \beta_{k}=1$, and the normalized covariance of noise is 1 for all pairs of channels: the noise is perfectly correlated.
We discuss these examples further in \S\ref{sec:examples_redux} below.

The power spectrum is real, so the noise must real as well; but the cross-power spectrum can be complex. 
In this case, the noise forms an elliptical Gaussian distribution in the complex plane, and 
characterization of the noise requires another variance.
This measures the degree of elongation of the distribution in the complex plane:
\begin{eqnarray}
\delta (\tilde r_k \tilde r_{\ell} ) 
&=& \langle \tilde r_{k} \tilde r_{\ell} \rangle_n -  \langle \tilde r_{k}\rangle_n\langle \tilde r_{\ell} \rangle_n \label{eq:spect_noise_rkrk}\\
&=& {{1}\over{N_{\nu}^2}} \sum_{p,q,r,s=0}^{N_{\nu}-1} 
\tilde g_k^{{x}}\, \tilde f_{k-p}^{{x}}\, 
\tilde g_{\ell}^{{y} *}\, \tilde f_{{\ell}-s}^{{y} *}\, 
\left\langle\tilde e_p^{{x}}\, \tilde e_s^{{y} *}\right\rangle_n
\tilde g_{\ell}^{{x} }\, \tilde f_{\ell-r}^{{x}}\, 
\tilde g_{k}^{{y} *}\, \tilde f_{k-q}^{{y} *}\, 
\left\langle\tilde e_r^{{x}}\, \tilde e_q^{{y} *}\right\rangle_n \nonumber \\
&=& \tilde \rho_k\, \tilde \rho_{\ell}\, \tilde \beta_{k-\ell}^{{xy}}\, \tilde \beta_{k-\ell}^{{xy} *} \cdot \delta_{e}. \nonumber 
\end{eqnarray}
Here, we have made use of the hybrid quantity
$\tilde \beta_{\ell}^{{xy}}$, the Fourier transform of $\beta^{{xy}}$ in Eq.\ \ref{eq:beta_def}.
Eqs.\ \ref{eq:spect_noise_rkrk} and\ \ref{eq:spect_noise_rkrkc} show that noise is strongest in phase with signal 
(\citealp[see][Eq.\ 20]{Gwi06}; \citealp[][\S 4]{Gwi11}).
If the underlying noiselike series $\tilde e_k^{{x}}$ and $\tilde e_k^{{y}}$ are uncorrelated,
$\delta (\tilde r_k \tilde r_{\ell} )=0$; in that case, the distribution of noise in the complex plane is a circular Gaussian distribution.

\subsection{Lag Domain: Correlation Function}\label{sec:lag_domain}

\subsubsection{Average correlation function}\label{sec:average_correlation_function}

Correlation characterizes the observed time series in the lag domain, Fourier conjugate to the spectral domain.
The cross-correlation function $r_{\kappa}$ is given by:
\begin{eqnarray}
r_{\kappa} &=& {{1}\over{N_{\nu}}} \sum_{\ell=0}^{N_{\nu}-1} x_{\ell} \; y_{\ell+\kappa}^* \label{eq:r_def}
\end{eqnarray}
We define the autocorrelation function $a_{\kappa}^{{x}}$ analogously, but of course by auto-correlation of identical time series.
We can determine the statistical averages through a direct average of Eq.\ \ref{eq:r_def},
or by using the convolution theorem for the Fourier transforms of $x_{\ell}$ and $y_{\ell}$ with Eq.\ \ref{eq:tilde_rho_as_g}.
We find
\begin{eqnarray}
\rho_{\kappa} \equiv \langle r_{\kappa}\rangle_n
&=& \sum_{\ell=0}^{N_{\nu}-1} g_{\ell}^{{x}}\, g_{\ell+\kappa}^{{y} *}\cdot \delta_{e} . \label{eq:rho_tau_as_g}  \\ 
\alpha_{\kappa}^{{x}} \equiv \langle a_{\kappa}^{{x}}\rangle_n &=&  \sum_{\ell=0}^{N_{\nu}-1} g_{\ell}^{{x}}\, g_{\ell+\kappa}^{{x} *}, \nonumber
\end{eqnarray}
and similarly for $\alpha_{\kappa}^{{y}}$.
Like the average spectrum,
the average correlation function depends only on the propagation kernel, not on the time-envelope function.
Note that our normalization condition (Eq.\ \ref{eq:g_normalization}) implies $\alpha_{0}=1$.
Again, if the underlying noiselike time series $e_{\ell}^{{x}}$ and $e_{\ell}^{{y}}$ are different, then $\rho_{\kappa}=0$.

\subsubsection{Noise in the correlation function}\label{sec:math_correlation_noise}

Now we consider the noise in the lag domain.
We can obtain this by the inverse Fourier transform of the noise in the spectral domain, Eq.\ \ref{eq:spect_noise_rkrkc}.
\begin{eqnarray}
\delta (r_\kappa r_\kappa^*) &=& \langle r_{\kappa} r_{\kappa}^* \rangle_n -  \langle r_{\kappa}\rangle_n\langle r_{\kappa}^* \rangle_n \label{eq:xcf_noise} \\
&=& {{1}\over{N_{\nu}^2}} \sum_{k,\ell=0}^{N_{\nu}-1} 
e^{-i {{2\pi}\over{N_{\nu}}} k \kappa}  e^{i {{2\pi}\over{N_{\nu}}} \ell \kappa}\,
\tilde \alpha_k^{{x}}\, \tilde \alpha_{\ell}^{{y} *}\, \tilde \beta_{k-\ell}^{{x}}\, \tilde \beta_{k-\ell}^{{y} *} \nonumber \\
&=& {{1}\over{N_{\nu}^2}} \sum_{\mu=0}^{N_{\nu}-1} 
\alpha_{\kappa-\mu}^{{x}}\, \alpha_{\kappa-\mu}^{{y} *}\; \sum_{\nu=0}^{N_{\nu}-1} \beta_{\nu}^{{x}}\, \beta_{\mu+\nu}^{{y}}\, . \nonumber
\end{eqnarray}
Thus, noise in the lag domain is the correlation of the time-envelopes at the two stations,
convolved with the product of the autocorrelation functions at the two stations.
Here, we have used Eq.\ \ref{eq:convolution_theorem_variation} to re-express $\tilde \beta_{k-\ell}\, \tilde\beta_{k-\ell}^*$,
and the facts that $\beta$ is real, and that $\alpha_{\kappa}=\alpha_{-\kappa}^*$.
The expression for the noise in the autocorrelation function, $\delta(a_{\kappa} a_{\kappa}^*)$, is identical, with ${y}\rightarrow {x}$.
As this equation shows,
in the lag domain, the noise in a single lag is affected by the time-envelope:
unlike signal in either domain, or noise in one channel of the spectral domain.
If the underlying noiselike series $e_{\ell}^{{x}}$, $e_{\ell}^{{y}}$ are distinct for $x$ and $y$, then the noise is still given by Eq.\ \ref{eq:xcf_noise}, unchanged.
For the power spectrum of the time series $x_{\ell}$, the same expression again gives the noise, with of course $r_{\kappa} \rightarrow a_{\kappa}^{{x}}$ on the left-had side of the equation, 
and with $\alpha_{\kappa}$ and $\beta_{\kappa}$ all for station $x$ on the right: ${y}\rightarrow {x}$. The analogous expressions hold for $y_{\ell}$.

The practitioner may be surprised that intermittent emission affects noise in the lag domain,
but not in the spectral domain.
After all, noise in one domain is the Fourier transform of noise in the other.
However, we calculate the variance of noise;
and correlations in one domain affect variances in the other, a consequence of the convolution theorem.
The order of imposition of time variation and convolution is important:
if the emission from the source were first convolved with a propagation kernel,
and then modulated in time, then the time-envelope would affect noise in the spectral domain.

In general, correlation is complex in the lag domain: $\rho_{\kappa}$ and $\alpha_{\kappa}$ are complex.
We require a second variance to fully characterize it. This can be found by Fourier transform of Eq.\ \ref{eq:spect_noise_rkrk}:
\begin{eqnarray}
\delta (r_\kappa r_\kappa) &=& \langle r_{\kappa} r_{\kappa} \rangle_n -  \langle r_{\kappa}\rangle_n\langle r_{\kappa} \rangle_n \label{eq:xcf_noise_rr} \\
&=& {{1}\over{N_{\nu}^2}} \sum_{k,\ell=0}^{N_{\nu}-1} 
e^{-i {{2\pi}\over{N_{\nu}}} k \kappa}  e^{-i {{2\pi}\over{N_{\nu}}} \ell \kappa}\,
\tilde \rho_k\, \tilde \rho_{\ell}\, \tilde \beta_{k-\ell}^{{xy}}\, \tilde \beta_{k-\ell}^{{xy} *} \cdot\delta_{e} \nonumber \\  
&=&
{{1}\over{N_{\nu}^2}} \sum_{k,\ell,\mu,\nu=0}^{N_{\nu}-1} 
e^{-i {{2\pi}\over{N_{\nu}}} (k+\ell) \kappa} \,
\tilde \rho_k\, \tilde \rho_{\ell}\, 
e^{-i {{2\pi}\over{N_{\nu}}} (k-\ell) \mu} \,
\beta_{\nu}^{{xy}}\, \beta_{\mu+\nu}^{{xy}} \cdot\delta_{e} \nonumber \\ 
&=& {{1}\over{N_{\nu}^2}} \sum_{\mu=0}^{N_{\nu}-1} 
\rho_{\kappa+\mu}\, \rho_{\kappa-\mu}\; \sum_{\nu=0}^{N_{\nu}-1} \beta_{\nu}^{{xy}}\, \beta_{\mu+\nu}^{{xy}}\cdot\delta_{e}  . \nonumber 
\end{eqnarray}
This gives the ellipticity of the distribution of noise in the complex plane.
Here, we have again used Eq.\ \ref{eq:convolution_theorem_variation}.
If the underlying noiselike series $e_{\ell}^{x}$, $e_{\ell}^{y}$ are distinct for $x$ and $y$, then $\delta (r_\kappa r_\kappa) = 0$,
and the distribution of noise is circular.
The expression for the noise in the autocorrelation function of the time series $x_{\ell}$ is again identical, with $r_{\kappa} \rightarrow a_{\kappa}^{{x}}$ and
$\rho \rightarrow \alpha$ as well as 
${y}\rightarrow {x}$.

\subsection{Examples}\label{sec:examples_redux}

In this section we present some simple examples of the theory above. 
In the examples, we suppose that $x_{\ell}$ and $y_{\ell}$ are identical,
so that our observations measure intensity for a single noise source. We therefore consider autocorrelation functions
$a_{\kappa}$ and spectrum $\tilde a_k$,
and their statistically-averaged counterparts $\alpha_{\kappa}$ and $\tilde \alpha_{k}$.

\subsubsection{Constant-intensity source}\label{sec:constant_intensity_source}

\begin{figure}[t]
\epsscale{.90}
\plotone{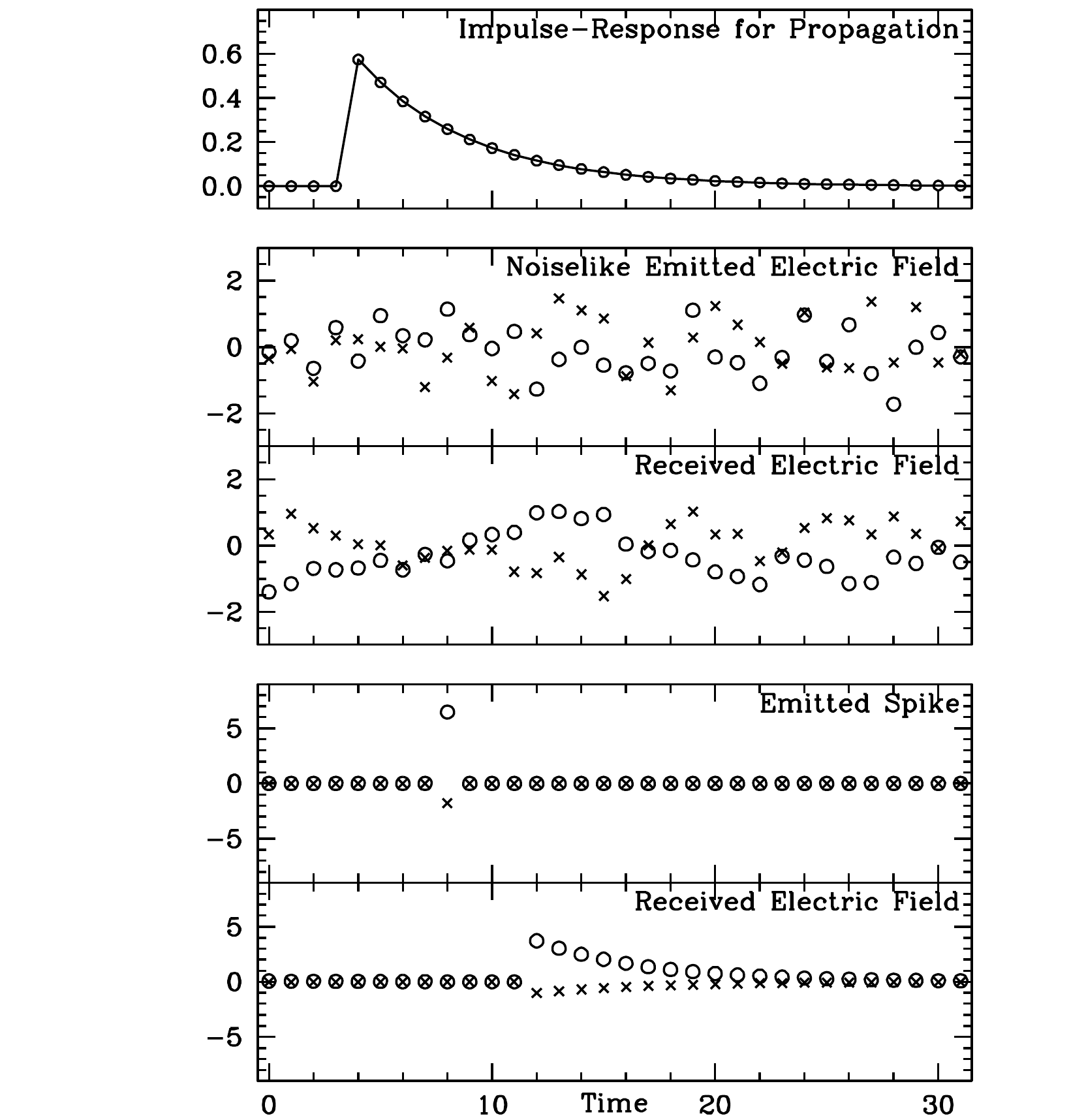}
\figcaption[]{
Effects of convolution with a propagation kernel (top panel) 
on a time series consisting of constant-intensity noiselike emission
(middle pair of panels),
and on a single spike (lower pair of panels).
Constant-intensity emission is drawn from a Gaussian distribution with unit average power.
The spike is drawn from the same distribution, but with amplitude adjusted so that the mean square power is the same in the two cases.
Circles show real parts, crosses imaginary parts.
\label{fig:simplesim_2.pdf}}
\end{figure}

In the simple and traditional case of constant source intensity over the integration time,
source electric field is simply
Gaussian white noise, with constant variance.
Such emission characterizes most continuum sources.
Propagation introduces correlations between samples at different times,
without changing the variance of the distribution.
Figure \ref{fig:simplesim_2.pdf} shows such a convolution,
for a short span of random electric field and a schematic propagation kernel.
The electric field is drawn from a complex Gaussian distribution at each sample.
Real and imaginary parts have variances of ${{1}\over{2}}$, so that the intensity is 1.
The propagation kernel is a one-sided exponential, $g_j ={{1}\over{T_s}} \exp(-t/T_s)$ for $t>0$, with characteristic time $T_s=5$.
This kernel satisfies the normalization condition for $g_j$, Eq.\ \ref{eq:g_normalization}.
This resembles effects of scattering,
with a rapid rise and slow fall,
although the complicated amplitude and phase variations are absent.
As the figure shows, the convolution smooths the electric field. Because the kernel is normalized, the intensity is unchanged.
From this time series, one can calculate a single realization of the autocorrelation function.
Alternatively, one can Fourier transform the electric field and form its square modulus: this is one realization of the spectrum.

\begin{figure}[t]
\epsscale{.90}
\plotone{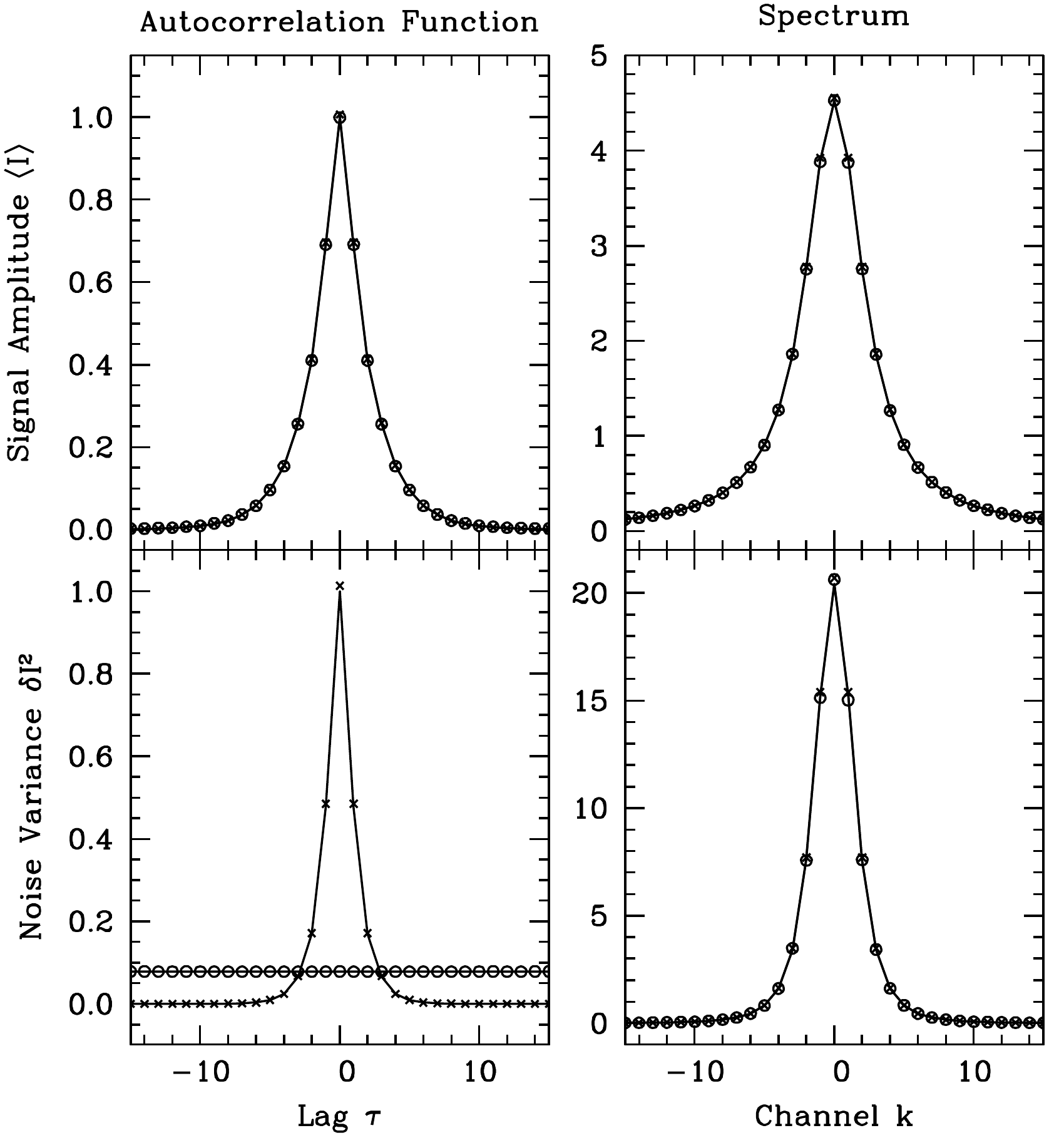}
\figcaption[]{
Correlation functions and spectra of constant-intensity noiselike emission and single-spike emission, convolved with a one-sided exponential.
Solid lines show expected forms, and circles (for constant intensity) and crosses (for spike) show signal and variance of noise for 
$10^5$ simulations like that in Figure\ \ref{fig:simplesim_2.pdf}.  
Upper panels: signals in correlation (left) and spectral (right) domains.
Lower panels: noise.  Left panel shows the correlation function, with flat curve for constant-intensity noiselike emission and sharp peak for a delta-function spike.
\label{fig:rwalk_noiseconv}}
\end{figure}

The average autocorrelation function and average spectrum agree with predictions of 
\S\ref{sec:avg_spectrum} and \S\ref{sec:average_correlation_function}.
Figure \ref{fig:rwalk_noiseconv} compares results of numerical simulation with the theoretical results.
For a source with white noiselike emission with constant intensity, the time-envelope is constant:
$f_{\ell} = 1$, and so $\beta_{\ell}=1$ and $\tilde\beta_k = \delta_{k\, 0}$.
The average autocorrelation function $\alpha_{\kappa}$ is simply the autocorrelation function of the propagation kernel $g_j$,
as Eq.\ \ref{eq:rho_tau_as_g} indicates. In this case, it is
a two-sided exponential, $\alpha_\kappa = \exp(-|\kappa|/T_s)$.
The average spectrum is the square modulus of the Fourier transform of $g_j$, as Eq.\ \ref{eq:tilde_rho_as_g} indicates:
a Lorentzian with half-width at half maximum $N_{\nu}/(2 \pi T_s)$, for our discrete Fourier transform.
In the examples in Figure\ \ref{fig:simplesim_2.pdf} and\ \ref{fig:rwalk_noiseconv},
$T_s$ approaches $N_{\nu}$, so the forms depart slightly from these analytic forms in our calculations and simulations.

In our model, noise arises purely from source noise. Noise agrees well with theoretical estimates, as Figure \ref{fig:rwalk_noiseconv} shows.
For a constant-intensity source,
the noise in one channel of the spectrum is equal to the square modulus of the average spectrum in that channel
(Eq.\ \ref{eq:spect_noise}, \citealp{Bro06}).
The signal-to-noise ratio is 1, but this can be increased by averaging together samples of the spectrum 
with the same propagation kernel.  
The noise of the autocorrelation function is constant as a function of lag.
Eq.\ \ref{eq:xcf_noise} shows that it is $\delta(a_{\kappa} a_{\kappa}^*) = {{1}\over{N_{\nu}}} \sum_{\nu=0}^{N_{\nu}-1} |\alpha_{\nu}|^2 = 2/T_s$. 
For the individual realizations, noise must be identical at positive and negative lags: this is a consequence of the fact that the power spectrum is
always real, so that $a_\kappa = a_{-\kappa}^*$, even for an single realization.

\subsubsection{Single-spike emission}\label{sec:spike_mathematical}

As an example from the opposite extreme, suppose that the time envelope consists of a single spike: a delta-function.
The time-envelope 
is $f_j = \sqrt{N_{\nu}}\, \delta_{j 0}$.
This satisfies Eq.\ \ref{eq:f_normalization}, the normalization condition for $f_j$.
The electric field at the source is nonzero only for one sample, with the random electric field of that sample drawn from a Gaussian distribution.
The observed time series is the 
convolution of the resulting spike with the propagation kernel, yielding a copy of the propagation kernel $g_j$,
scaled by the random electric field of the spike, and shifted in time.
The observed time series is a single, one-sided exponential, with random amplitude and phase.
Figure\ \ref{fig:simplesim_2.pdf} shows the resulting time series, and effects of the propagation kernel. 

For our propagation kernel, 
the autocorrelation function is a two-sided exponential, times a random gain factor $|e_0|^2$; and the spectrum its Fourier transform.
However, averaging over many such realizations recovers the same averages as in the constant-intensity 
example, as Figure\ \ref{fig:simplesim_2.pdf} shows.

Noise is quite different.
The noise is a single gain-like factor, which changes the amplitude of the spectrum but does not change its form.
Noise in the spectrum turns out to be simply the square of the average spectrum,
just as in the case of white noise.
However, each realization of the spectrum is a copy of the average spectrum, scaled by a random factor,
so that noise is perfectly correlated among channels.
Similarly, in the lag domain, each realization of the autocorrelation 
function is a copy of the average, scaled by some amount.
Consequently, the variance of noise in the lag domain is the square of the autocorrelation function.
The distribution of noise is not uniform, as for a constant-intensity source.
Noise is concentrated near zero lag, and falls rapidly away from it.
Figure\ \ref{fig:rwalk_noiseconv} shows this example as well.

\subsubsection{Multiple spikes}\label{sec:random_spiky_acf}

Emission in multiple, randomly spaced spikes
is intermediate between continuous emission and a single spike.
\citet{Cor76} proposed such shot noise as a fundamental model for pulsar emission.
We suppose that the pulsar emits $n_e$ single-sample spikes during the observing period of $N_{\nu}$ samples,
and that emission is zero at other times.
We suppose that all spikes have equal amplitude, and that they are randomly spaced.
Then $f_p = \sqrt{N_{\nu}/n_e}\ {\rm or}\ 0$. 
We consider a propagation kernel with phase as well as amplitude variations.

Figure\ \ref{fig:rwalk_noiseconv_59} shows an example.
The source emits 6 spikes of equal average intensity.
The time-envelope $f_{\ell}$ is shown at the bottom of the top panel of the figure.
Each spike has a random electric field, drawn from a Gaussian distribution.
The spikes are then dispersed by a propagation kernel $g_{\ell}$: a complex sinusoid with a  declining envelope.
We suppose that $x$ and $y$ are identical: we measure $a_{\kappa}$ and $\tilde a_{k}$.
We average results over many realizations with different noise, but with identical time-envelope and propagation kernel.
The averages are computed with spikes at identical locations (that is, $f_{\ell}$ is fixed), but varying source noise $e_{\ell}$.
As the upper panel shows, convolution with the kernel $g$ disperses the spikes in time, and allows them to interfere if they overlap.
The form of the kernel is apparent.
The average correlation function in the middle panel
shows the imprint of phase variations in the propagation kernel, and of its characteristic decline with time.  
Any effect of correlation between different spikes is absent in the average, because differences in phase of 
$e_{\ell}$ between spikes cancel it.
The correlation function displays the required normalization $\alpha_0 = 1$ and symmetry $\alpha_{\kappa} = \alpha_{-\kappa}^*$.

Correlation between different spikes contributes to noise, as shown in the lowest panel.
Noise in the correlation function is largest at the central lag, but nonzero to the highest lags.
Vertical lines at the bottom of the lower panel show the autocorrelation of the time envelope, $\beta_k$.
The distribution of noise displays the overall behavior of $\beta_{\kappa}$. 
The spiky $\beta_{\kappa}$ is smoothed by convolution with the square of the average correlation function 
$\alpha_{\kappa}$.  
This leads to a relatively smooth ``floor'' level of noise for all $\kappa$.
This ``floor'' arises from Eq.\ \ref{eq:xcf_noise}, and is the same for real and imaginary parts.
Eq.\ \ref{eq:xcf_noise_rr} dictates dependence of the noise on phase,
and is nonzero only where individual convolved spikes interact with themselves via the terms $\rho_{\kappa+\mu}$ and $\rho_{\kappa-\mu}$.
This takes place at $\kappa\rightarrow 0$ and $\kappa\rightarrow N_{\nu}/2$.
Consequently, the noise depends on phase in these regions.
The ``bump'' at large positive and negative lag arises from the latter.
If desired, it can be removed by zero-padding the time series.

\subsubsection{Random emission events}\label{sec:ACF_cartoon}

In general, we expect spikes, or other patterns of emission, to vary in location among different samples of the spectrum.  They may also 
vary in number and strength,
and have finite width and various shapes.
Description of these effects requires averaging over the time-envelope $f_{\ell}$.
These effects can be included in our model by dividing a much longer span of observations into segments,
and finding effects of time integration by convolution of the spectrum.
Here, we briefly discuss the form of such spectra, and effects of averaging
over time-envelope as well as noise.

\begin{figure}[t]
\epsscale{.90}
\plotone{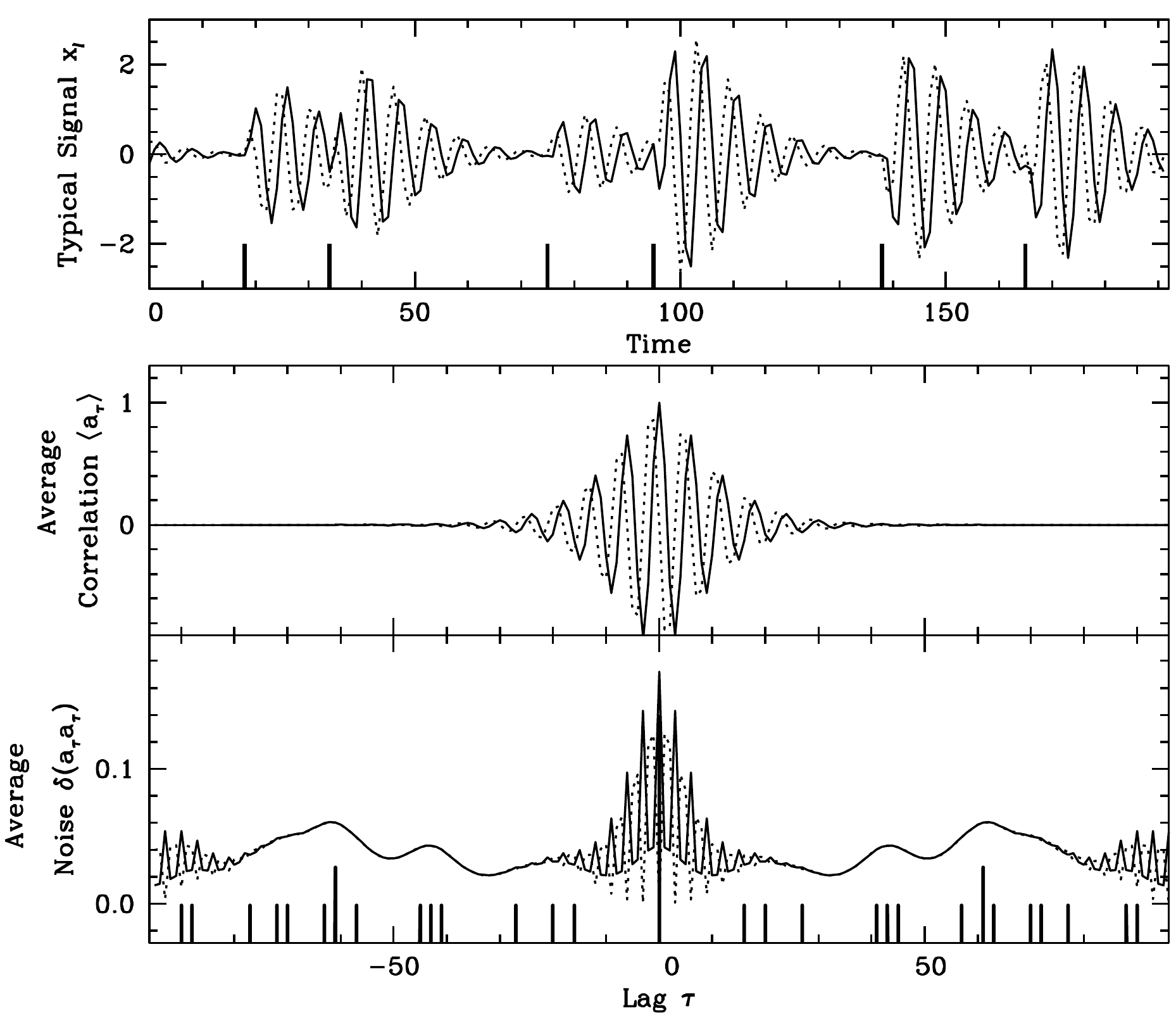}
\figcaption[]{
Autocorrelation function and noise for random spiky emission and more complicated propagation kernel.
Top: Typical time series.  Vertical lines at lower edge show 6 spikes of the time-envelope $f_p$, displaced downward.
Solid line shows real part, and dotted line shows imaginary part, of one time series $x_{\ell}$,
including effects of multiplication by random $e_{\ell}$ and convolution with 
propagation kernel $g_p$.
Middle: Average correlation function for time series, with $f_p$ and $g_p$
as in top panel. 
Lower: Noise for correlation function. 
Solid line shows variance of real part, dotted line shows variance of imaginary part.
Vertical lines at lower edge show $\beta_\kappa$, displaced downward. 
\label{fig:rwalk_noiseconv_59}}
\end{figure}

Because the average correlation function $\alpha_{\kappa}$ is affected only by the propagation kernel $g_{\ell}$,
variations of the time-envelope $f_{\ell}$ do not affect it.
They affect only the noise, through the autocorrelation function of $\beta_{\kappa}$ in the second sum in Eq.\ \ref{eq:xcf_noise}.
Suppose that the emission consists of a number $n_e$ of
randomly-placed spikes of width one sample, with uniform variance, as above. 
Then $\beta_{\kappa}=N_{\nu}/n_e\ {\rm or}\ 0$.
The correlation function of
$\beta_{\kappa}$, times $1/N_{\nu}^2$, 
is a peak of area $N_{\nu}^2$ at $\kappa=0$,
and $n_e (n_e - 1)$ peaks of area $N_{\nu}^2/n_e^2$ distributed elsewhere.
The central peak results from correlation of each spike with itself, and the other peaks from correlation of spikes at different times.
Convolution with the square of $\alpha_{\kappa}$
smooths the spiky correlation function of $\beta_{\kappa}$ to a more nearly flat ``floor'' level, 
as Figure\ \ref{fig:rwalk_noiseconv_59} illustrates.
The area of the central peak is the product of areas of the convolved functions,
$A=\sum_{\kappa} |\alpha_{\kappa}|^2$, and area of each of the outlying peaks is $A /n_e^2$.

Averaging over many spectra with the same number of spikes at different locations, and the same propagation kernel,
leaves the central peak unchanged, while smoothing the noise floor to a constant level.
The central peak contains noise power $A$, and the floor contains total power $A (n_e-1)/n_e$,
or power per lag of $A (n_e-1)/(n_e N_{\nu} )$.
For a single emission event $n_e\rightarrow 1$, the peak dominates and the noise floor vanishes, as expected  
(see \S\ref{sec:spike_mathematical}).
For many spikes $n_e>> N_{\nu}$, the noise floor dominates and the noise is distributed approximately uniformly in lag, as expected for stationary noiselike emission
(see \S\ref{sec:constant_intensity_source}).
As \citet{Cor76} notes, superposition of sufficiently many spikes results in stationary noise.

If the emission takes the form of ``events'' of some typical width $w_e$,
then the resulting average autocorrelation function takes the same form:
a central, broadened peak and a surrounding noise floor.
The normalization of $\beta_{\kappa}$ demands that the central peak have unit area (although now width $\sim w_e$),
and that the outlying peaks again each have area $N_{\nu}/n_e$ (and width $\sim w_e$).  
Noise in the autocorrelation function takes a similar form:
a central peak of noise power $A$, and a uniform noise floor of power per lag $A (n_e-1)/(n_e N_{\nu} )$.

\begin{figure}[t]
\epsscale{.90}
\plotone{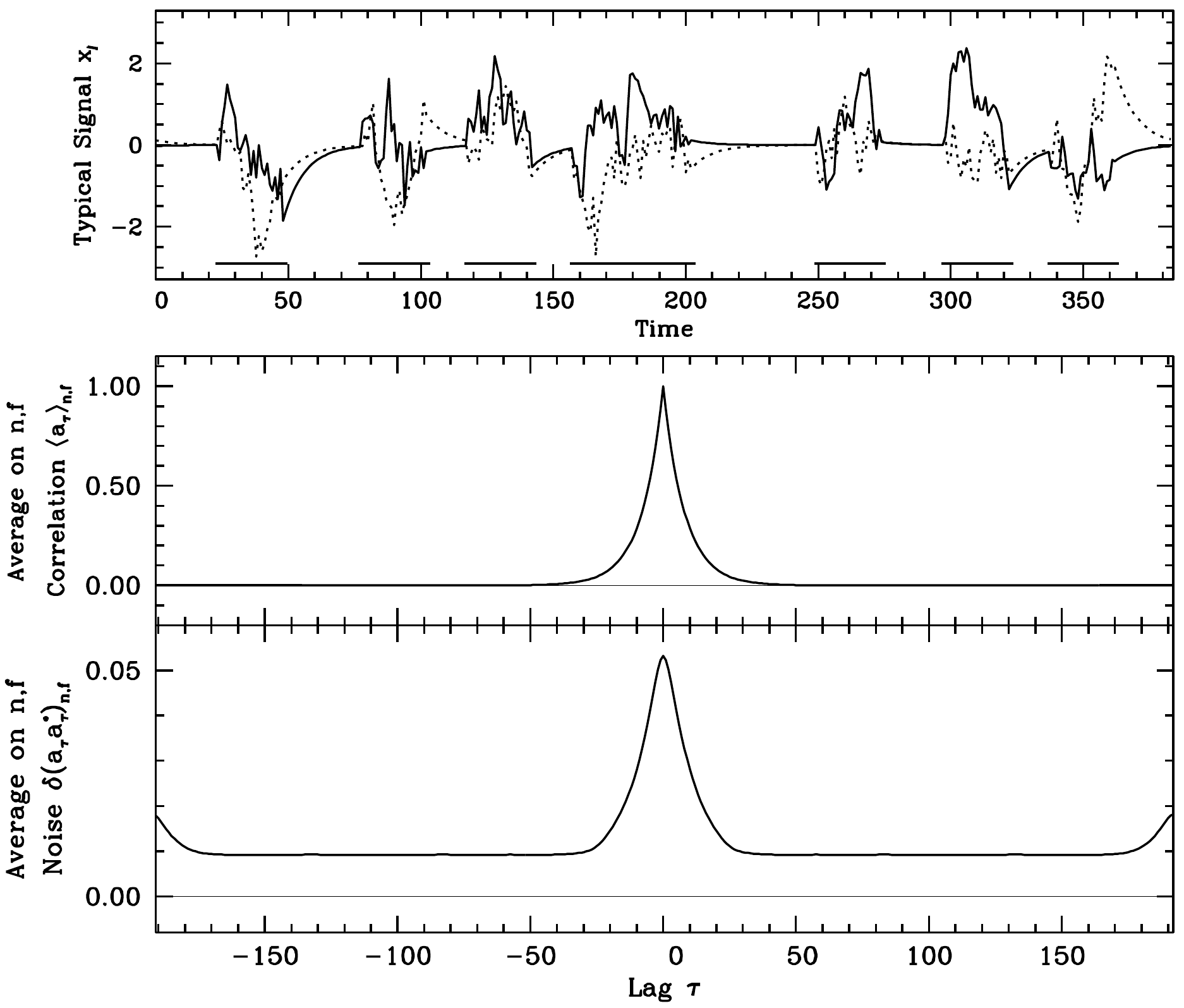}
\figcaption[]{
Autocorrelation function and noise for wider bursts of source emission, with varying times of bursts.
Top: Typical example of time series.  Horizontal lines at lower edge show time-envelope of emission events, where $f_p>0$.
During each of 8 bursts, width $w=13$, source emission is noiselike with constant variance,
and is convolved with the one-sided exponential propagation kernel.
Solid line shows real part, and dotted line shows imaginary part, of one typical time series $x_{\ell}$.
Middle: Average correlation function, for average over source noise and over times of emission events. 
Lower: Variance of noise for correlation function. 
The offset from zero is the noise ``floor'' described in the text.
Noise peak is broadened, relative to the peak in the middle panel, by the finite width of the bursts.
\label{fig:ACF_cartoon}}
\end{figure}

Figure\ \ref{fig:ACF_cartoon} shows an example of random emission events:
the average autocorrelation function and noise for a time series containing
8 randomly-placed time-envelopes of length 13 and constant height, and an exponentially-declining real propagation kernel.
The average includes random placement of the events, as well as varying noise.
Signal reflects only the propagation kernel; noise reflects propagation kernel and the time-envelope.
The noise shows a floor of the expected level. The central noise peak is broadened by the autocorrelation 
function of the individual emission events.
In contrast to results for 1 or more spikes in
Figures\ \ref{fig:rwalk_noiseconv} and \ref{fig:rwalk_noiseconv_59}, the noise peak is broader than the signal peak.
This is because of the greater width of the emission events $w_e$, and consequent broadening of $\beta_{\kappa}$.

\section{DISCUSSION}\label{sec:extensions}

\subsection{Connection with Amplitude-Modulated Noise Theory}\label{sec:amplitude_modulated_noise}

Our model for pulsar emission is based on the amplitude-modulated noise theory of \citet{Ric75}.
This work
provides a test of whether pulsar emission is stationary Gaussian noise,
multiplied by a variable time envelope \citep{Ric75}.
This model
has become a standard benchmark for studies of statistics of pulsar emission \citep{Jen01,Smi03,Pop09,Jes10}.
Both our and Rickett's analyses are concerned with the second and fourth moments of the electric field.
As a test of this property, \citet{Ric75} invokes
the correlation function of intensity $R_I(\kappa) = \langle I(t) I(t+\kappa)\rangle_{t,n}$.
Here, the angular brackets include both an average over time, and over realizations of noise, as indicated by the subscripts.
This function is related to the statistics described in the present paper,
particularly to noise, the fourth moment of electric field.
Rickett includes in his analysis 
a convolution in time, similar to our propagation kernel, to describe effects of a bandpass filter in the receiver. 
However, he does not include propagation effects explicitly.
\citet{Smi03} point out that propagation effects can be difficult to remove, and residual ones can strongly affect the 
observed form of $R_I$.

Our treatment differs from \citet{Ric75} in that we include propagation effects explicitly,
including those with timescale $\tau$ longer than the timescale of variation of the source.  
Rickett assumes that the minimum timescale for variation of the time-envelope 
(his $w$, our $w_e$ for width of an emission event) must be much 
longer than timescale associated with the kernel (his $1/B$ for the inverse of the receiver bandwidth, our $\tau$ for the propagation kernel). 
For pulsar B0834$+$06, for example, this demands that flux density variations of the pulsar be longer than 1.2\ msec,
much longer than many of the variations that are observed.
Our expressions in \S\ref{sec:math_treatment} hold for arbitrary time-envelope and propagation kernel;
of course, our kernel can include receiver response as well.
We also include both auto- and cross-correlation, and we distinguish between signal and noise contributions.
We do not invoke an average over time, as Rickett does, but only over the statistics of the noiselike electric field of the source.

Rickett's paper and the present work have fundamentally different aims: 
Rickett proposes a test of whether a source emits amplitude-modulated Gaussian noise in the absence of propagation effects,
whereas we calculate effects of propagation under the assumption of amplitude-modulated noise.
It would be interesting to test whether the source emission is noiselike, in the presence of propagation effects.
This requires additional knowledge of, or assumptions about, propagation and source emission.
In a promising approach, \citet{Jes10} compare the form of $R_I$ for different pulse phases; specifically, for giant pulses 
from the Crab pulsar.
Interstellar propagation is not likely to produce differences between pulse phases.
Comparison of signal and noise for the autocorrelation functions, for different pulse phases for this object, 
would produce observables that we could compare with the present work.

\subsection{Background Noise}\label{sec:with_background_noise}

Background noise often dominates self-noise and is always present.
Background noise is not coherent with source noise.
Thus, we can take the electric field as a superposition of time series $x^s_{\ell}$ for the source and $x^n_{\ell}$ for noise,
and calculate the signal and noise for each pair of fields separately.
To maintain generality, we must relax the assumption that the time series have unit variance.
We assign variances $I_s$ and $I_n$;
these are the average flux densities of of signal and noise, respectively.
We maintain the convention that $\tilde\alpha$ and $\tilde\rho$ are normalized as above, and adopt $I_s$ and $I_n$ as prefactors, as discussed in \S\ref{sec:flux_density}.
For interferometry, background noise is uncorrelated between antennas, and we must distinguish between backgrounds at the two
antennas $I_{nA}$ and $I_{nB}$. 
Usually background noise is of constant intensity, so that $\tilde \alpha^n_k=1$ for all $k$; generalization to spectrally-varying background noise is straightforward.
In the spectral domain, we recover the expressions for interferometric visibility $\tilde V_k$:
\begin{eqnarray}
\langle \tilde V_k \rangle_n &=&  \left( I_s \tilde \rho^s_k \right) \label{eq:noise_polynomial_visibility}\\
\delta (\tilde V_k \tilde V^*_k )&=& 
 \left( I_s \tilde \alpha^{s}_k\right)^2
+  \left( I_s \tilde \alpha^{s}_k \right) \left\{ (I_{nA}+I_{nB})\right\} +\left\{ I_{nA} I_{nB}\right\}  \nonumber \\
\delta (\tilde V_k \tilde V_k )&=& 
\left( I_s \tilde \rho^{s}_k\right)^2 . \nonumber
\end{eqnarray}
For a relatively short baseline, $\tilde \rho_{k}\approx \tilde \alpha_k$,
and the expressions for variance become a second-order polynomial in 
the average visibilty $\langle \tilde V_k \rangle_n$ for the real part,
and a first-order polynomial for the imaginary part \citep[see][Eq.\ 8]{Gwi11}.
Similarly, for single-dish observations, we obtain in the spectral domain the intensity $\tilde I_k$:
\begin{eqnarray}
\langle \tilde I_k \rangle_n &=& I_s \tilde\alpha^{s}_k + I_n \label{eq:noise_polynomial_intensity}\\
\delta (\tilde I_k^2 )&=& 
 \left( I_s \tilde \alpha^{s}_k\right)^2+  \left( I_s \tilde \alpha^{s}_k \right) \left\{2 I_{n}\right\} + \left\{ I_{n}^2 \right\}. \nonumber
\end{eqnarray}
Again, the expression for noise is a second-order polynomial in the average intensity \citep[see][Eq.\ 11]{Gwi11}.
The polynomial is a perfect square, for intensity.
When averaged over $N_t$ samples in time, with constant flux density in each spectral channel, 
each term declines by a factor of $1/N_t$;
the polynomial remains a perfect square.

If the average flux density of the source changes slowly as spectra are averaged,
the polynomial coefficients can decline more slowly than $1/N_t$.
We present a detailed model in \citet{Gwi11}, \S 2.2.2.
In the simplest cases, such as a pulsar with a rectangular pulse,
the effect can be treated as a reduction in the number of independent samples.
For more complicated amplitude variations,
the coefficients depend on the moments of the distribution of mean intensity, and the
polynomial is no longer a perfect square.

\subsection{Example Observations}\label{sec:conclude_example}

We propose that variations of $f$ on timescales shorter than the accumulation of a single spectrum explain the apparent
lack of noise at large lags for pulsar B0834$+$06 \citep{Gwi11}.
The distribution of noise in the spectrum is in agreement with that for a constant-intensity source, 
following the polynomial forms given by Eqs.\ \ref{eq:noise_polynomial_visibility} and \ref{eq:noise_polynomial_intensity},
with corrections for amplitude variations \citep[][\S 2.2.2]{Gwi11}.
This noise, must be conserved in a transform to the correlation function,
in accord with Parseval's Theorem.
In this domain, the noise is uniformly distributed for interferometric observations,
which are dominated by background noise.
However, noise appears deficient at large lags of the correlation function for single-dish observations,
which are dominated by self-noise.
We suggest that this apparent deficiency is precisely the concentration of noise at small lags predicted for an intermittent source,
as seen in Figures\ \ref{fig:rwalk_noiseconv_59} and \ref{fig:ACF_cartoon}
and discussed in \S\ref{sec:ACF_cartoon}.

We can form a quantitative estimate of the degree of source intermittency using the results of \S\ref{sec:ACF_cartoon}.
In the single-dish observations, 
we measure a noise level of about 40\% of that expected for stationary noiselike emission, from noise
in the spectral domain and Parseval's Theorem,
at lags greater than $40\ \mu{\rm sec}$.
About a quarter of this is
sky or instrumental noise, which is uniformly distributed.
Thus, we suggest that self-noise is reduced to about 30\% of its constant-intensity value at large lags, by intermittency.

In terms of the model from \S\ref{sec:ACF_cartoon},
reduction to 30\% 
indicates about 1.5 emission events per integration period of $300\ \mu{\rm sec}$.
The inferred duration of the events is $w_e<40\ \mu{\rm sec}$.
Peaks of emission of this short duration, repeating at about this rate, are indeed observed for this pulsar 
\citep{Kar78}.
Of course, alternative models, for example involving occasional powerful narrow emission events interspersed with more typical 
constant emission,
can explain the observations equally well.

\section{SUMMARY}\label{summary}

In this paper, we consider the effects of intermittent emission at the source
on observations of individual realizations of the spectrum and correlation function.
We argue in \S\ref{sec:convolution} that effects of propagation can be described as a convolution in time, or multiplication in frequency, to high accuracy.
We assume that the underlying emission at the source is 
white noise drawn from a stationary Gaussian distribution, modulated by a time-envelope function $f(t)$, 
as 
discussed in \S\ref{sec:noise_def} and \S\ref{sec:math_treatment}.
We suppose that this emission is convolved with a propagation kernel $g$, and that the resulting electric field 
is correlated at the observer, as described mathematically in \S\ref{sec:correlation}.

We exploit the properties of Gaussian noise to 
find that variations of the time-envelope do not affect the average spectrum, the average correlation function,
or the noise in individual channels in the spectrum, in \S\ref{sec:math_spectrum} and \ref{sec:average_correlation_function}.
These are all determined only by the propagation kernel and the flux density of the source.
The time-envelope does affect noise in the correlation function,
and the correlation of noise between different channels in the spectrum. We present expressions for the distribution of signal and noise,
and spectral correlations of noise.  Table\ \ref{equation_table} summarizes our mathematical results.

The distribution of noise in the correlation function depends on the form of the intermittency.
For an source of constant intensity, the noise is constant with lag (\S\ref{sec:constant_intensity_source}).
For a single spike, the emission is zero outside a central peak (\S\ref{sec:spike_mathematical}).
For multiple randomly-occurring emission events, the noise has a constant ``floor'' surrounding a central peak (\S\ref{sec:ACF_cartoon}).
The width of the peak is set by the timescale of the propagation kernel and the widths of the events.
We contrast our calculation with the amplitude-modulated noise theory of \citet{Ric75}, which does not include propagation effects.

Background noise adds a constant offset, and a linear term, to the polynomial describing noise as a function of source intensity (\S\ref{sec:with_background_noise}).
We apply our theory to recent observations of noise in the secondary spectrum of the scintillating pulsar B0834$+$06,
and show that these suggest that the emission of the pulsar is intermittent on timescales shorter than $40\ \mu{\rm sec}$ (\S\ref{sec:conclude_example}).
Our theory suggests, interestingly, that the existence of intermittency can be inferred even when temporal smearing,
via convolution with the propagation kernel, has reduced the degree of modulation.  Additional work, now in progress,
suggests that more extensive analysis can recover both the propagation kernel and the form of the emission, at least in statistical form.

\acknowledgments

We thank the National Science Foundation (AST-1008865) for financial support.

\newpage

\newpage

\newpage

\clearpage
\begin{deluxetable}{llll}
\tablecolumns{4}
\tablewidth{0pc}
\tablecaption{Dictionary of Symbols}
\tablehead{
\colhead{Symbol}&\colhead{Definition}&\colhead{Reference} &\colhead{Notes}\\
}
\startdata
\sidehead{Time Series}
$x_{\ell}$ & Observed Electric Field & Eq. \ref{eq:x_ell_def} & a,b \\
$e_{\ell}^{{x}}$ & Underlying Stationary Noise& \S\ref{sec:noise} & a,b \\
$f_{\ell}^{{x}}$ & Time-Envelope Function& \S\ref{sec:time_envelope} & a,b \\
$g_{\ell}^{{x}}$ & Propagation Kernel& \S\ref{sec:propagation} & a,b  \\
$\beta_{\ell}^{{x}}$ & Square Modulus of $f_{\ell}$ & Eq.\ \ref{eq:beta_def} & a,b  \\
\sidehead{Spectral Domain}
$\tilde a_k^{{x}} $ & Power Spectrum & Eq.\ \ref{eq:single_realization_ac_spectrum} & \\  
$\tilde\alpha_k^{{x}}$ & Average Power Spectrum & Eq.\ \ref{eq:tilde_alpha_as_g} & a \\ 
$\delta (\tilde a_k \tilde a_k)$& Noise in Power Spectrum &\S\ref{sec:math_spectrum_noise} & \\ 
$\tilde r_k^{{x}} $ & Cross-power Spectrum & Eq.\ \ref{eq:single_realization_spectrum} & \\
$\tilde\rho_k$ & Average Cross-Power Spectrum & Eq.\ \ref{eq:tilde_rho_as_g} & \\
$\delta (\tilde r_k \tilde r_k^*)$,  $\delta (\tilde r_k \tilde r_k)$& Noise in Cross-Power Spectrum & Eqs.\ \ref{eq:spect_noise_rkrkc}, \ref{eq:spect_noise_rkrk} & \\
$I$ & Intensity & & \\
$V$ & Interferometric Visibility & & \\
$I_s$ & Average Flux Density of Source & \S\ref{sec:flux_density}, \S\ref{sec:with_background_noise} & \\
$I_n$ & Flux Density of Noise & \S\ref{sec:with_background_noise} & \\
\sidehead{Lag Domain}
$a_{\kappa}$ & Autocorrelation Function &\S\ref{sec:average_correlation_function} &\\
$\alpha_{\kappa}^{{x}}$ & Average Autocorrelation Function& Eq.\ \ref{eq:r_def} & a \\
$\delta (a_{\kappa} a_{\kappa}^*)$, $\delta (a_{\kappa} a_{\kappa})$& Noise in Autocorrelation Function & \S\ref{sec:math_correlation_noise}  & \\
$r_{\kappa}$ & Cross-correlation Function & Eq.\ \ref{eq:r_def} &\\
$\rho_{\kappa}$ & Average Cross-correlation Function & Eq.\ \ref{eq:rho_tau_as_g} & \\
$\delta (r_{\kappa} r_{\kappa}^*)$, $\delta (r_{\kappa} r_{\kappa})$& Noise in Cross-correlation Function & Eqs.\ \ref{eq:xcf_noise}, \ref{eq:xcf_noise_rr} & \\
\enddata
\tablenotetext{a} {For quantities for companion field $y_{\ell}$, substitute $x\rightarrow y$. }
\tablenotetext{b} {Fourier transforms of these quantities indicated by accent $\ensuremath{\sim}$.}
\label{symbol_table}
\end{deluxetable}

\clearpage
\begin{deluxetable}{lll}
\tablecolumns{3}
\tablewidth{0pc}
\tablecaption{Summary of Equations}
\tablehead{
\colhead{Description}&\colhead{Equation$^{a}$}&\colhead{Reference} \\
}
\startdata
\sidehead{Time-Envelope Function}
Normalization of $f$              & $\sum_{\ell} f_{\ell}^{{x}} f_{\ell}^{{x}} = N_{\nu} $ &Eq. \ref{eq:f_normalization}  \\
Normalization of $\tilde f$      &$\sum_{k} \tilde f_k^{{x}} \tilde f_k^{{x} *} =N_{\nu}$ & \S\ref{sec:data_series_in_spectral_domain} \\
Definition of $\beta$                & $\beta_\ell^{{x}} = f_{\ell}^{{x}} f_{\ell}^{{x}}                   $ &Eq.  \ref{eq:beta_def}              \\
Normalization of $\beta$         & $\sum_{\ell}  \beta_\ell^{{x}} = N_{\nu}             $ & see Eq. \ref{eq:f_normalization}  \\
Normalization of $\tilde\beta$  & $\tilde \beta_{0}^{{x}} = \tilde \beta_{0}^{{y}} = 1$ & \S\ref{sec:math_spectrum_noise} \\
\sidehead{Propagation Kernel}
Normalization of $g$         & $\sum_{\ell}\, g_\ell^{{x}}\, g_\ell^{{x}*} = 1$      & Eq. \ref{eq:g_normalization} \\
Normalization of $\tilde g$ & $\sum_{k} \tilde g_k^{{x}} \tilde g_k^{{x} *} =1$ & \S\ref{sec:data_series_in_spectral_domain} \\
\sidehead{Spectrum}
Average Cross-Power Spectrum$^{b}$   & $\tilde \rho_k =  \tilde g_{k}^{{x}}\, \tilde g_{k}^{{y} *}\cdot \delta_{e}$ & Eq. \ref{eq:tilde_rho_as_g} \\ 
Average Power Spectrum & $\tilde \alpha_k^{{x}} = \tilde g_{k}^{{x}}\, \tilde g_{k}^{{x} *}$ & Eq.\ \ref{eq:tilde_alpha_as_g}\\ 
Normalization of $\tilde \alpha$               & $\sum_{k} \tilde\alpha_k = 1$ & see \S\ref{sec:data_series_in_spectral_domain} \\
Noise in Single Channel$^{c}$ & $\delta (\tilde r_k \tilde r_k^* )= \tilde \alpha_k^{{x}}\, \tilde \alpha_k^{{y} *}$& Eq. \ref{eq:spect_noise} \\
Covariance of Noise in 2 Channels$^{c}$ & 
$\delta (\tilde r_k \tilde r_{\ell}^* ) =  \tilde \alpha_k^{{x}}\, \tilde \alpha_{\ell}^{{y} *}\, \tilde \beta_{k-\ell}^{{x}}\, \tilde \beta_{k-\ell}^{{y} *} $ & Eq.\ \ref{eq:spect_noise_rkrkc} \\
Conjugate $^{b,d}$& $\delta (\tilde r_k \tilde r_{\ell} ) =  \tilde \rho_k\, \tilde \rho_{\ell}\, \tilde \beta_{k-\ell}^{{xy}}\, \tilde \beta_{k-\ell}^{{xy} *}\cdot \delta_{e}$ & Eq. \ref{eq:spect_noise_rkrk} \\ 
\sidehead{Correlation  Function}
Average Cross-Correlation Function$^{b}$ & $\rho_{\kappa} =  \sum_{\ell} g_{\ell}^{{x}}\, g_{\ell+\kappa}^{{y} *} \cdot \delta_{e}$ & Eq.\ \ref{eq:rho_tau_as_g} \\ 
Average Autocorrelation Function & $\alpha_{\kappa}^{{x}} =  \sum_{\ell} g_{\ell}^{{x}}\, g_{\ell+\kappa}^{{x} *}$ & Eq.\ \ref{eq:rho_tau_as_g} \\
Normalization of $\alpha_{\kappa}$ & $\alpha_0 = 1$ & \S\ref{sec:average_correlation_function} \\
Noise in a Single Lag$^{c}$ & $\delta (r_\kappa r_\kappa^*)= {{1}\over{N_{\nu}^2}} \sum_{\mu} 
\alpha_{\kappa-\mu}^{{x}}\, \alpha_{\kappa-\mu}^{{y} *}\; \sum_{\nu} \beta_{\nu}^{{x}}\, \beta_{\mu+\nu}^{{y}}$ & Eq.\ \ref{eq:xcf_noise} \\
Conjugate$^{b,c}$ &$\delta (r_\kappa r_\kappa) = {{1}\over{N_{\nu}^2}} \sum_{\mu} 
\rho_{\kappa+\mu}\, \rho_{\kappa-\mu}\; \sum_{\nu} \beta_{\nu}^{{xy}}\, \beta_{\mu+\nu}^{{xy}}\cdot \delta_{e}$ & Eq.\ \ref{eq:xcf_noise_rr} \\ 
\enddata
\tablenotetext{a} {All sums run from 0 to $N_{\nu}-1$.}
\tablenotetext{b} {If noise series $e_x$, $e_y$ are identical, then $\delta_{e}=1$; if noise series are different, then $\delta_{e}=0$.}
\tablenotetext{c} {For expressions for noise in power spectrum $\tilde a_k^{{x}}$ or autocorrelation function $a_{\kappa}^{{x}}$, substitute $y\rightarrow x$, $\rho\rightarrow\alpha$. } 
\tablenotetext{d} {Vanishes for autocorrelation function $\alpha$ and power spectrum $\tilde\alpha$. } 
\label{equation_table}
\end{deluxetable}

\end{document}